\documentclass[final]{pasj01}

\Received{$\langle$reception date$\rangle$}

\Accepted{$\langle$acception date$\rangle$}

\Published{$\langle$publication date$\rangle$}

\usepackage{graphicx}
\usepackage{rotating}
\usepackage{afterpage}
\usepackage{environ}
\NewEnviron{myresizeenv}{\resizebox{\linewidth}{!}{\BODY}}
\usepackage[driver=dvips]{expl3}
\usepackage{txfonts}
\usepackage[sort&compress]{natbib}
\bibpunct{(}{)}{;}{a}{}{,}
\newcommand{\um}{$\mu$m}

\def\arcsec{\hbox{$^{\prime\prime}$}}
\def\utw{\smash{\rlap{\lower5pt\hbox{$\sim$}}}}
\def\udtw{\smash{\rlap{\lower6pt\hbox{$\approx$}}}}

\def\farcs{\hbox{$.\!\!^{\prime\prime}$}}
\def\farcm{\hbox{$.\!\!^{\prime}$}}

\def\Msun{\hbox{\it M$_\odot$}}

\def\Mbol{\hbox{\it M$_{bol}$}}

\def\K{\hbox{\it K}}

\def\Mk{\hbox{\it M$_{\rm K}$}}

\newcommand{\Ks}{{\it K$_{\rm s}$}}

\newcommand{\Aks}{{\it A$_{\it K_{\rm s}}$}}

\def\simgr{\mathrel{\hbox{\rlap{\hbox{\lower4pt\hbox{$\sim$}}}\hbox{$>$}}}}
\begin{document}

\title{ANDICAM $I$ and $J$-band monitoring of bright inner Galactic late-type stars.}
\author{Maria~Messineo\altaffilmark{1,2}
}

\altaffiltext{1}{Key Laboratory for Researches in Galaxies and Cosmology, University of 
Science and Technology of China, Chinese Academy of Sciences, Hefei, Anhui, 230026, China}
\altaffiltext{2}{ ASTROMAGIC freelancer,  Potsdam 14482, Germany}


\maketitle

\begin{abstract}
Time-series photometry in $I$- and $J$-band of 57 inner Galactic 
late-type stars, highly-probable red supergiant (RSG) stars, 
is here presented.
38\% of the sample presents significant photometric variations.
The variations in  $I$ and $J$band appear to be correlated,
with $\Delta I \propto \Delta J \times 2.2$, 
$\Delta I$ variations ranging from 0.04-1.08 mag,
$\Delta J$ variations from 0.03-0.52 mag.
New short periods ($<1000$ d) could be estimated for 8 stars and
range from 167-433 d.
This work confirms that the sample is not contaminated by
large-amplitude Asymptotic Giant Branch (AGB) stars.
Furthermore, despite the large errors in distance,
the period-luminosity diagram suggests that the sample is 
populating the same sequence as the known Galactic RSGs.

\end{abstract}
%\linenumbers

\section{Introduction}

Red supergiant (RSG) stars are late-type stars
burning helium in the central core and with initial masses
larger than 8-9 \Msun. They are intrinsically bright 
at infrared wavelengths and, in principle, they can be detected 
at great distances even in the most obscured
regions of the Galaxy. Unfortunately, 
the  resemblance between RSGs and  
Asymptotic Giant Branch  (AGB) stars
and the lack of precise distances complicate their detection.

In the last decade, the astronomical community has made 
a great effort to conduct 
medium- and high-resolution spectroscopic studies of RSGs. 
Nowadays, metallicity and temperatures can  be directly 
inferred from iron lines \citep[e.g.,][]{taniguchi21}, 
and some infrared lines have been found with  strengths 
that correlate with 
the stellar luminosity \citep[e.g.,][]{messineo21}.
The spectroscopic future  looks promising with 
millions of spectra to be released by the 
Gaia,  LAMOST,  GALAH, and 4MOST, surveys 
\citep[e.g.,][]{deJong12,gaia21, wu21,  sharma22}.
This will allow us to greatly improve the census of Galactic RSGs
and to better learn how to classify them well and 
at  minimum cost.

For improved Galactic distances, one must wait for the new releases 
of Gaia parallaxes and the pulsational periods  of long-period 
variable stars. Periods for millions of stars  will  soon be available 
from the Gaia survey and the forthcoming LSST survey 
\citep{ivezic19}. 
Periods may yield distance estimates via a period-luminosity relation.\\ 
It has long been established  that
there is a correlation between the length of the period
and the stellar luminosity of variable late-type stars.
That such  a relation also exists  for variable RSG stars
was noticed by \citet{glass79} 
by analyzing seven RSGs in the LMC.
The work in the LMC  by \citet{feast80} confirmed it with 
the analysis of 24 RSGs analyzed.
The RSG  absolute \K\ magnitudes (\Mk) versus Periods (Per) 
relation falls above that
of AGB stars, and their $K$-band amplitudes are typically
smaller than 0.25 mag, while in AGB stars amplitudes 
range from 0.5-1.0 mag \citep{wood83}.
Starting around the end of the 90s, the description of 
the long-period variables became more complicated
with the discovery of several parallel sequences of pulsators  
in the LMC \citep[][]{ita04} and multi-frequencies
detected in their light curves \citep[e.g.,][]{soszynski13}.
In the M 33 galaxy, \citet{soraisam18} detected
a  well defined period-luminosity relation for RSGs 
pulsating in the fundamental mode, and a parallel sequence,  
likely, of first-overtone pulsators  0.3 mag brighter.

In the Milky Way, \citet{pierce00} report that
 12  RSGs in the Perseus OB1 association
follow the same period-luminosity relations determined for 
RSGs in the LMC and  M 33 in $R,I,K$-bands.
\citet{kiss06} analyze the  light curves of about 40 Galactic 
M-type RSGs covering 61 years and found that
about 40\% of them have two periods, a short period  
($<1000$ d)  and a long secondary period
(LSP) greater than 1000 d.
The same sample of Galactic stars and others from the LMC and M 33
(220 stars) were analyzed by \citet{chatys19}, to find
that, for variable RSGs, a period-luminosity exists 
for the short periods and it appears to be universal
and independent of metallicity. About 52\% of the Galactic RSG
sample has short periods and 47\% long periods.

In the last decade, a large number of new Galactic 
RSGs have been reported
in the literature \citep[e.g.,][]{messineo17, dorda18}.
At the current time, in view of the  forthcoming surveys,
it is advisable to aim for a more precise and well-established
characterization of already detected objects, as those will
constitute the reference frames for the new stars to be classified.
Especially for those located towards the densest
and obscured regions of the Milky Way, variability
studies are of primary importance
to confirm their nature,  estimate their distance, and, 
therefore, their luminosity class.

\section{The sample}
The 57 late-type stars observed with ANDICAM are 
taken from the sample of 94 stars analyzed 
by \citet{messineo16} and \citet{messineo17}.

The sample of \citet{messineo16} comprises 
stars from the GLIMPSE I North survey  brighter than \Ks = 7 mag  
and  with \Aks $>$ 0.4 mag, which satisfies the  
infrared color criteria advised by 
\citet[][]{messineo12}.
\citet{messineo17} show that large equivalent widths of the CO at 2.29 \um\ 
(EW(CO) $>45$ \AA) and lack of water vapour absorption
featured 62\% of that sample.
The sample is, therefore, mostly made  up 
of red supergiants (RSGs)
\citep[see discussion in][]{messineo17}.

The subsample observed with ANDICAM is listed in Table \ref{table1},
and comprises 55 stars with broad EW(CO) ($>45$ \AA) and 
little water, which \citet{messineo17} label as ``EW$>$'', plus
two other stars (MZM7 and MZM21), which are labeled as ``CO'' by  
having EW(CO) larger than 37 \AA.

Unfortunately, as  described in the recent work of
\citet{messineo21}\footnote{There is no overlap between 
the ANDICAM sources and the sources in \citet{messineo21}}, 
the EW(CO)  alone is not a good 
luminosity indicator and  distance estimates and variability information 
remain essential to determine the luminosity class.

In the following, the stars are called late-type stars, 
as their classification is based 
on the EW of the CO band-heads at 2.29  \um\ from low-resolution spectra.
A more solid classification can be foreseen with high-resolution spectra. However,
the broadness of the EW(CO), together with the lack of water absorption, current distance 
uncertainty, and the small amplitudes here presented, 
suggest that they are all consistent with being RSGs.

\subsection{Infrared photometry and distances}
The collection of infrared photometric 
measurements and bolometric magnitudes
are presented in the works
of \citet{messineo17} and \citet{messineo19}
and listed in Table  \ref{table1}.

In \citet{messineo16}, distances  are determined 
by comparing the target extinction with the extinction curves
of nearby clump stars, which are primary indicators of distance.
In \citet{messineo19}, Gaia DR2 parallaxes are matched 
to the infrared sources and used to infer their distances.
A revised version of this Gaia catalog with EDR3
parallaxes was made available by \citet{messineo21z}.
Kinematic distances  using the Gaia DR2 velocities
are only possible for seven sources. 
The absolute magnitudes in \Ks, \Mk,
are calculated as \Ks-\Aks-DM,
where Ks are the 2MASS Ks magnitudes
\Aks is the interstellar extinction, which is derived from the observed  $H-Ks$ and $J-Ks$, 
by assuming the intrinsic colors of 
\citet{koornneef83} and the extinction coefficients 
by \citet{messineo05}, which assumes an infrared power 
law with an index of $-1.9$.  
DM is the distance moduli from the Gaia EDR3 parallaxes.

\begin{figure*}
\begin{center}
\resizebox{0.48\hsize}{!}{\includegraphics[angle=0]{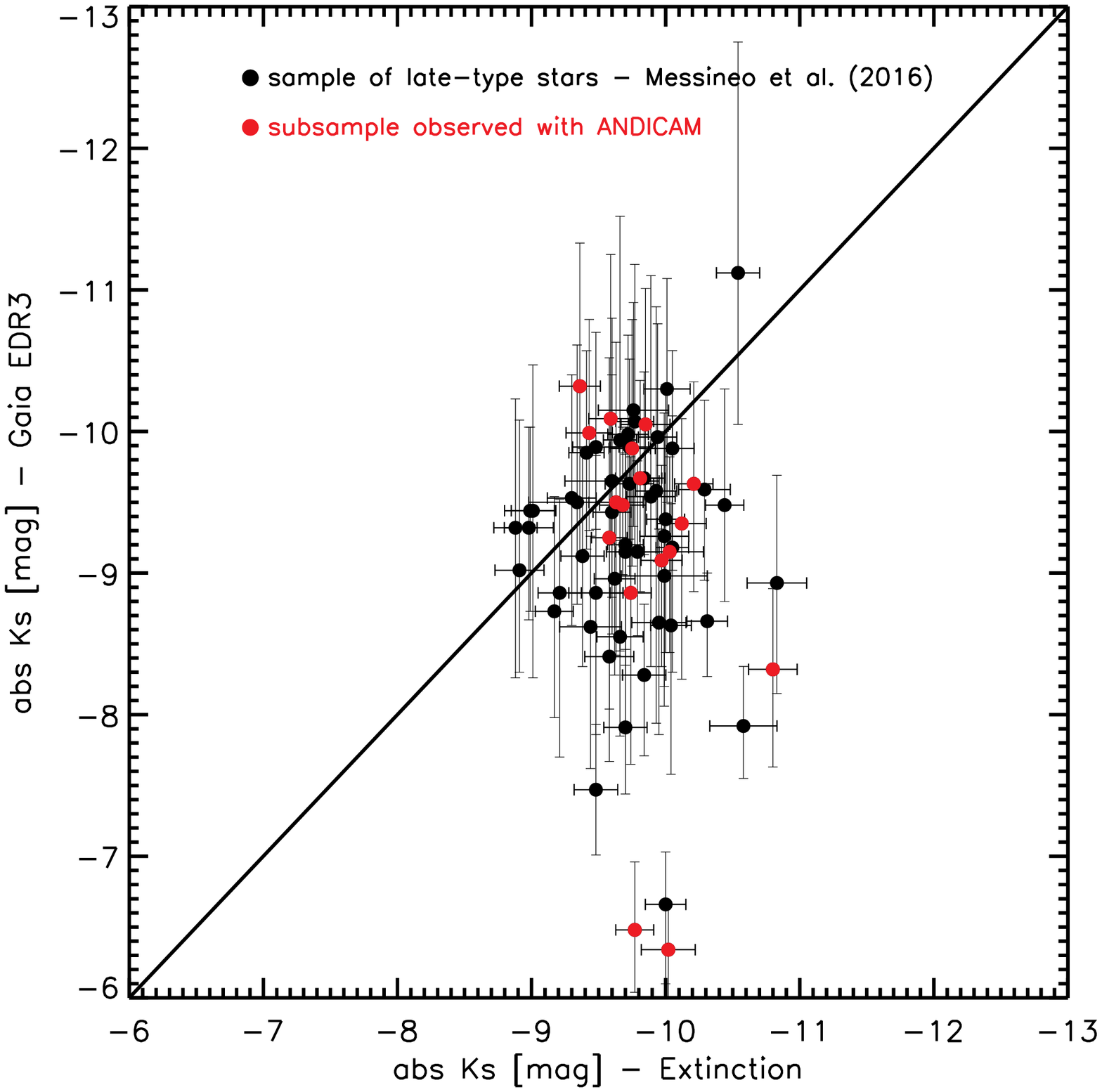}}
\resizebox{0.48\hsize}{!}{\includegraphics[angle=0]{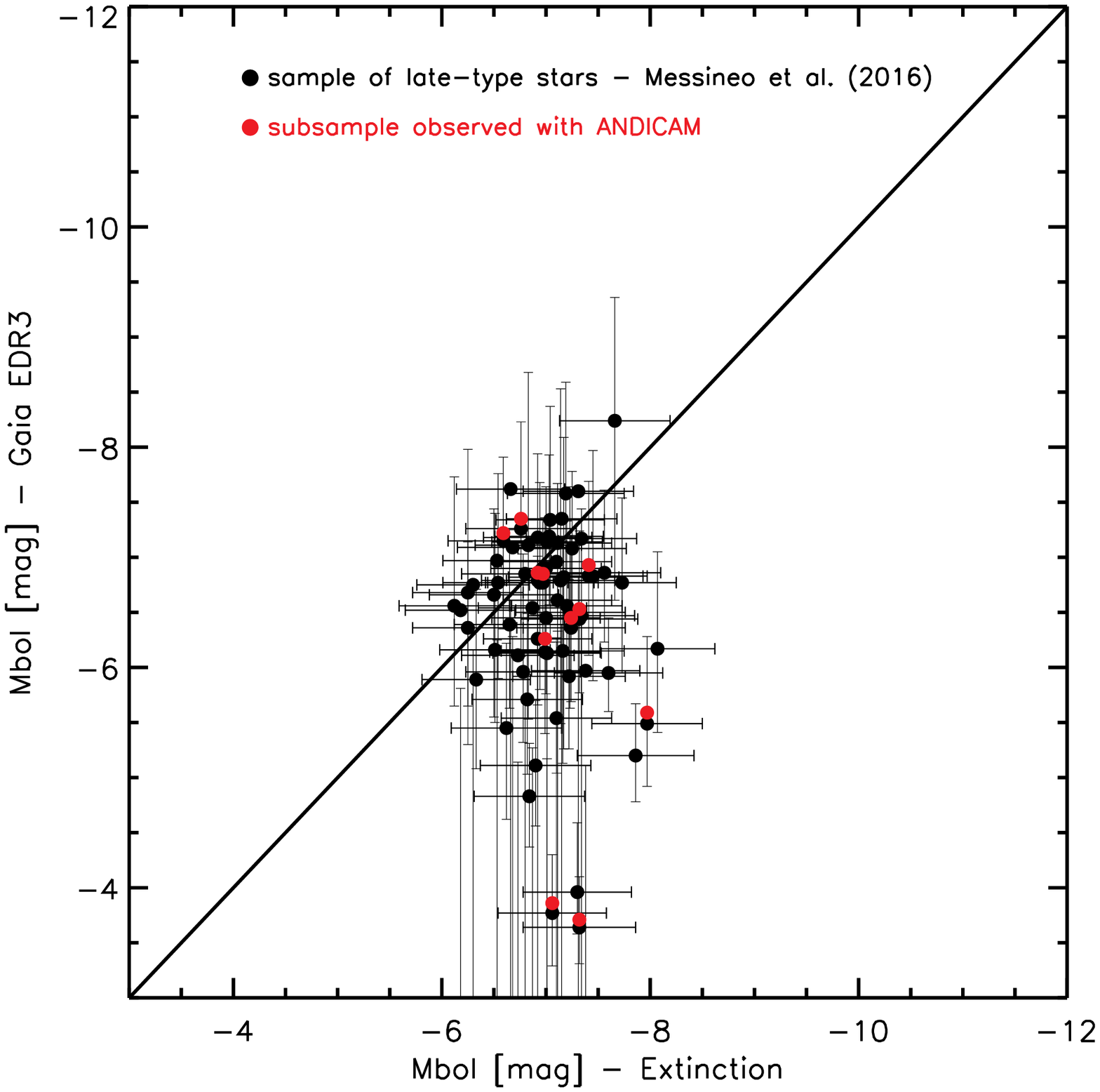}}
\end{center}
\caption{\label{figdist} {\it Right panel:} 
Sample of highly-likely  M-type RSGs reported by \citet{messineo16} 
and \citet{messineo17}. 
The  subsample  observed with ANDICAM is marked in red.
The absolute \Ks\ magnitudes, \Mk, calculated with Gaia EDR3 parallaxes
\citep{messineo21z} are plotted vs. those  based on 
extinction calculations \citep{messineo16,messineo17}. 
{\it Left panel:} The \Mbol\ values calculated with Gaia EDR3 parallaxes
\citep{messineo21z} are plotted vs. those  based on 
extinction calculations \citep{messineo16,messineo17}. } 
\end{figure*}

Unfortunately, fractional parallactic errors are large
for this type of cool sources and distances, with
resulting errors in the distance moduli mostly between 
0.8 and 1.0 mag. However, a comparison with the values inferred
using the extinction is  interesting
because it further confirms the source location in the inner Galaxy,
as shown in Fig.\ \ref{figdist}. 
The final Gaia release will narrow down these errors.

\begin{table*}
\vspace*{-.5cm} 
\caption{\label{table1} List of observed late-type stars and  parameters from literature.}
\begin{myresizeenv}
\begin{centering}
{\tiny
\renewcommand{\arraystretch}{0.7}
\begin{tabular}{llllllrllllll}
%\begin{tabular}{@{\extracolsep{-.06in}}llllrrrrrrrrrrrrrrrrrrrr}
\hline
\hline
2MASS-ID        &   MZM-ID&   RSG-Sp  &  \Aks &  DM                         & RUWE    & frac   & DM2      &DM3      &\Ks$_o$                     & Ampl(1) & Ampl(2) & Per(2) \\
                &         &           & [mag] & [mag]                       &         & [\%]   & [mag]    &[mag]    & [mag]                      & [mag]   & [mag]   & [d] \\
\hline
18112728-175014 &      MZM4 &      M0.5 &   1.38 &  $12.71 ^{ 0.88}_{-0.93 }$ &    2.1  &    0.4 &  $.. $ &  $13.58 ^{ 0.12}_{ 0.12 }$  &  $ 3.53  \pm 0.02    $  & 0.08  &$..$ &    $..$  \\
18114736-192915 &      MZM5 &        M2 &   1.66 &  $13.22 ^{ 1.64}_{-1.30 }$ &    1.3  &    0.2 &  $.. $ &  $13.57 ^{ 0.15}_{ 0.15 }$  &  $ 3.64  \pm 0.03    $  &$..$ &$..$ &    $..$  \\
18131562-180122 &      MZM6 &        M1 &   1.13 &  $13.81 ^{ 0.83}_{-0.91 }$ & $..$  &   -0.9 &  $.. $ &  $13.68 ^{ 0.17}_{ 0.17 }$  &  $ 3.93  \pm 0.03    $  &$..$ &$..$ &    $..$  \\
18132341-185818 &      MZM7 &        K4 &   1.54 &  $13.07 ^{ 0.92}_{-0.65 }$ &    1.1  &    1.2 &  $.. $ &  $13.57 ^{ 0.13}_{ 0.13 }$  &  $ 3.87  \pm 0.03    $  &$..$ &$..$ &    $..$  \\
18134914-184633 &      MZM9 &        M0 &   1.27 &  $12.90 ^{ 0.80}_{-0.69 }$ &    1.2  &    1.0 &  $.. $ &  $13.45 ^{ 0.14}_{ 0.14 }$  &  $ 3.75  \pm 0.02    $  &$..$ &$..$ &    $..$  \\
18140682-190620 &     MZM10 &      M0.5 &   1.10 &  $13.51 ^{ 0.80}_{-0.69 }$ & $..$  &    0.8 &  $.. $ &  $13.65 ^{ 0.18}_{ 0.18 }$  &  $ 3.84  \pm 0.03    $  & 0.09  &$..$ &    $..$  \\
18144593-173754 &     MZM11 &        M1 &   1.88 &  $ 8.68 ^{ 0.95}_{-0.64 }$ &    2.8  &    2.7 &  $.. $ &  $.. $  &  $ 3.43  \pm 0.03    $  &$..$ &$..$ &    $..$  \\
18154112-164645 &     MZM16 &      M1.5 &   1.05 &  $12.06 ^{ 0.73}_{-0.62 }$ &    1.1  &    2.2 &  $.. $ &  $.. $  &  $ 4.00  \pm 0.02    $  &$..$ &$..$ &    $..$  \\
18155832-165827 &     MZM17 &      M0.5 &   1.54 &  $13.30 ^{ 1.08}_{-1.13 }$ &    1.3  &    0.7 &  $.. $ &  $13.43 ^{ 0.20}_{ 0.20 }$  &  $ 3.80  \pm 0.03    $  &$..$ &$..$ &    $..$  \\
18171598-140554 &     MZM20 &      M0.5 &   0.91 &  $13.34 ^{ 1.20}_{-1.56 }$ &    1.2  &    2.1 &  $.. $ &  $13.69 ^{ 0.18}_{ 0.18 }$  &  $ 3.80  \pm 0.02    $  &$..$ &$..$ &    $..$  \\
18172865-163739 &     MZM21 &      K5.5 &   1.29 &  $12.92 ^{ 0.72}_{-1.06 }$ &    1.7  &    0.6 &  $.. $ &  $12.81 ^{ 0.18}_{ 0.18 }$  &  $ 3.90  \pm 0.03    $  &$..$ &$..$ &    $..$  \\
18174160-135628 &     MZM22 &        M0 &   1.11 &  $ 9.95 ^{ 0.38}_{-0.32 }$ &    1.4  &    4.5 &  $.. $ &  $13.63 ^{ 0.20}_{ 0.20 }$  &  $ 3.61  \pm 0.02    $  &$..$ &$..$ &    $..$  \\
18175212-171508 &     MZM23 &        M2 &   1.97 &  $13.39 ^{ 0.82}_{-1.15 }$ & $..$  &   -1.5 &  $.. $ &  $13.34 ^{ 0.35}_{ 0.35 }$  &  $ 3.74  \pm 0.03    $  &$..$ &$..$ &    $..$  \\
18184453-165108 &     MZM25 &        M0 &   2.41 &  $.. $ & $..$  &    0.0 &  $.. $ &  $.. $  &  $ 3.42  \pm 0.02    $  &$..$ &$..$ &    $..$  \\
18184660-163456 &     MZM26 &        M3 &   1.59 &  $12.98 ^{ 0.72}_{-1.11 }$ &    1.4  &    1.2 &  $.. $ &  $12.82 ^{ 0.36}_{ 0.36 }$  &  $ 3.48  \pm 0.03    $  &$..$ &$..$ &    $..$  \\
18210685-150340 &     MZM33 &      M0.5 &   1.21 &  $13.17 ^{ 1.21}_{-1.27 }$ & $..$  &    0.6 &  $.. $ &  $13.50 ^{ 0.13}_{ 0.13 }$  &  $ 3.92  \pm 0.03    $  & 0.09  &$..$ &    $..$  \\
18210846-153209 &     MZM34 &        M5 &   1.16 &  $12.18 ^{ 0.74}_{-0.83 }$ &    1.5  &    1.2 &  $.. $ &  $13.35 ^{ 0.18}_{ 0.18 }$  &  $ 3.77  \pm 0.03    $  & 0.09  &$..$ &    $..$  \\
18212427-135528 &     MZM35 &      M0.5 &   1.00 &  $10.27 ^{ 0.44}_{-0.48 }$ &    1.3  &    3.4 &  $.. $ &  $13.56 ^{ 0.14}_{ 0.14 }$  &  $ 3.79  \pm 0.02    $  &$..$ &$..$ &    $..$  \\
18230411-135416 &     MZM37 &        M2 &   1.08 &  $13.25 ^{ 0.76}_{-0.72 }$ &    1.4  &    0.4 &  $.. $ &  $13.83 ^{ 0.14}_{ 0.14 }$  &  $ 3.62  \pm 0.03    $  &$..$ &$..$ &    $..$  \\
18231119-135758 &     MZM38 &        M1 &   1.21 &  $13.06 ^{ 0.64}_{-0.63 }$ &    1.2  &    0.7 &  $.. $ &  $13.76 ^{ 0.19}_{ 0.19 }$  &  $ 3.47  \pm 0.03    $  &$..$ &$..$ &    $..$  \\
18240991-110121 &     MZM39 &      K5.5 &   0.74 &  $11.52 ^{ 0.52}_{-0.57 }$ &    1.3  &    2.8 &  $.. $ &  $.. $  &  $ 3.91  \pm 0.03    $  &$..$ &$..$ &    $..$  \\
18251120-114056 &     MZM40 &        M2 &   0.77 &  $11.72 ^{ 0.47}_{-0.55 }$ &    1.1  &    2.0 &  $.. $ &  $13.51 ^{ 0.16}_{ 0.16 }$  &  $ 3.81  \pm 0.02    $  & 0.15  & 0.52  &    $..$  \\
18254382-115336 &     MZM41 &        M0 &   0.71 &  $13.80 ^{ 0.65}_{-0.62 }$ &    0.9  &    1.2 &  $12.60 ^{ 0.21}_{-0.26 }$ &  $13.64 ^{ 0.14}_{ 0.14 }$  &  $ 3.91  \pm 0.03    $  &$..$ &$..$ &    $..$  \\
18261922-140648 &     MZM42 &        M3 &   0.74 &  $13.97 ^{ 0.69}_{-0.80 }$ &    1.1  &   -0.3 &  $.. $ &  $13.41 ^{ 0.17}_{ 0.17 }$  &  $ 3.98  \pm 0.03    $  & 0.09  &$..$ &    $..$  \\
18291233-121940 &     MZM45 &        M0 &   0.94 &  $12.77 ^{ 1.16}_{-0.66 }$ &    1.3  &    1.2 &  $11.48 ^{ 0.46}_{-0.68 }$ &  $13.12 ^{ 0.16}_{ 0.16 }$  &  $ 3.91  \pm 0.02    $  &$..$ &$..$ &    $..$  \\
18301889-102000 &     MZM46 &        M0 &   0.77 &  $14.21 ^{ 1.21}_{-1.01 }$ & $..$  &   -0.4 &  $.. $ &  $13.25 ^{ 0.15}_{ 0.15 }$  &  $ 3.89  \pm 0.03    $  & 0.07  & 0.33  &   215.37   \\
18310460-105426 &     MZM47 &        M1 &   1.07 &  $13.94 ^{ 0.93}_{-1.11 }$ &    1.4  &    0.2 &  $.. $ &  $13.64 ^{ 0.14}_{ 0.14 }$  &  $ 3.87  \pm 0.02    $  & 0.06  &$..$ &    $..$  \\
18315881-111921 &     MZM48 &        M0 &   0.96 &  $14.03 ^{ 0.80}_{-0.78 }$ & $..$  &   -0.6 &  $.. $ &  $13.74 ^{ 0.17}_{ 0.17 }$  &  $ 3.73  \pm 0.03    $  & 0.09  &$..$ &    $..$  \\
18334444-065947 &     MZM50 &        M1 &   0.85 &  $12.09 ^{ 0.57}_{-0.50 }$ &    0.9  &    2.4 &  $.. $ &  $13.65 ^{ 0.16}_{ 0.16 }$  &  $ 3.81  \pm 0.02    $  & 0.09  & 0.59  &    $..$  \\
18352902-072112 &     MZM54 &      M2.5 &   1.75 &  $13.89 ^{ 0.88}_{-1.16 }$ &    1.3  &    0.4 &  $.. $ &  $13.39 ^{ 0.16}_{ 0.16 }$  &  $ 3.80  \pm 0.03    $  &$..$ &$..$ &    $..$  \\
18353475-075648 &     MZM55 &        M0 &   1.60 &  $13.57 ^{ 0.79}_{-0.96 }$ & $..$  &   -0.5 &  $.. $ &  $13.37 ^{ 0.18}_{ 0.18 }$  &  $ 3.52  \pm 0.03    $  &$..$ &$..$ &    $..$  \\
18354911-073443 &     MZM56 &        M1 &   1.48 &  $12.18 ^{ 1.21}_{-0.96 }$ & $..$  &    1.9 &  $.. $ &  $13.06 ^{ 0.15}_{ 0.15 }$  &  $ 3.32  \pm 0.03    $  &$..$ &$..$ &    $..$  \\
18355151-073011 &     MZM57 &      M3.5 &   2.16 &  $13.73 ^{ 1.07}_{-1.63 }$ &    1.3  &    0.2 &  $.. $ &  $13.15 ^{ 0.16}_{ 0.16 }$  &  $ 2.61  \pm 0.02    $  &$..$ &$..$ &    $..$  \\
18374651-071224 &     MZM58 &      M1.5 &   1.71 &  $13.97 ^{ 1.03}_{-1.12 }$ & $..$  &   -0.6 &  $.. $ &  $.. $  &  $ 3.91  \pm 0.03    $  & 0.06  &$..$ &    $..$  \\
18413481-044857 &     MZM59 &      M1.5 &   1.05 &  $12.86 ^{ 1.00}_{-0.97 }$ & $..$  &    2.1 &  $.. $ &  $13.48 ^{ 0.20}_{ 0.20 }$  &  $ 4.00  \pm 0.03    $  &$..$ &$..$ &    $..$  \\
18414834-044852 &     MZM60 &      M1.5 &   1.14 &  $13.25 ^{ 1.06}_{-0.87 }$ &    1.9  &    0.4 &  $.. $ &  $13.98 ^{ 0.18}_{ 0.18 }$  &  $ 3.99  \pm 0.03    $  &$..$ &$..$ &    $..$  \\
18421710-044116 &     MZM61 &      M1.5 &   2.59 &  $.. $ & $..$  &    0.0 &  $.. $ &  $.. $  &  $ 3.38  \pm 0.02    $  &$..$ &$..$ &    $..$  \\
18424231-044053 &     MZM62 &        M3 &   0.90 &  $12.40 ^{ 0.70}_{-0.68 }$ &    1.3  &    1.5 &  $.. $ &  $13.51 ^{ 0.17}_{ 0.17 }$  &  $ 3.85  \pm 0.03    $  &$..$ & 0.29  &   109.56   \\
18424479-043357 &     MZM63 &      M0.5 &   0.81 &  $13.30 ^{ 0.49}_{-0.55 }$ &    1.1  &    0.8 &  $13.25 ^{ 0.14}_{-0.15 }$ &  $13.50 ^{ 0.17}_{ 0.17 }$  &  $ 3.82  \pm 0.03    $  &$..$ &$..$ &    $..$  \\
18425222-034618 &     MZM64 &      M2.5 &   1.56 &  $12.66 ^{ 0.92}_{-1.02 }$ &    1.4  &    0.5 &  $.. $ &  $13.67 ^{ 0.32}_{ 0.32 }$  &  $ 3.68  \pm 0.02    $  &$..$ &$..$ &    $..$  \\
18430800-035624 &     MZM65 &      M2.5 &   2.30 &  $.. $ & $..$  &    0.0 &  $.. $ &  $.. $  &  $ 4.04  \pm 0.03    $  &$..$ &$..$ &    $..$  \\
18442616-033527 &     MZM66 &      M0.5 &   1.35 &  $13.60 ^{ 0.86}_{-0.69 }$ & $..$  &   -1.1 &  $.. $ &  $13.77 ^{ 0.16}_{ 0.16 }$  &  $ 3.72  \pm 0.03    $  &$..$ &$..$ &    $..$  \\
18464441-032404 &     MZM67 &        M4 &   1.68 &  $10.36 ^{ 0.75}_{-0.56 }$ &    1.8  &    2.2 &  $.. $ &  $.. $  &  $ 3.35  \pm 0.02    $  &$..$ &$..$ &    $..$  \\
18464480-030332 &     MZM68 &        M2 &   1.16 &  $13.58 ^{ 0.81}_{-0.70 }$ & $..$  &   -0.4 &  $.. $ &  $13.32 ^{ 0.15}_{ 0.15 }$  &  $ 3.60  \pm 0.02    $  &$..$ &$..$ &    $..$  \\
18475357-014715 &     MZM69 &        M2 &   1.40 &  $13.79 ^{ 0.58}_{-0.81 }$ & $..$  &   -1.1 &  $.. $ &  $13.38 ^{ 0.17}_{ 0.17 }$  &  $ 3.90  \pm 0.03    $  &$..$ &$..$ &    $..$  \\
18482997-021150 &     MZM70 &      M1.5 &   1.25 &  $13.08 ^{ 0.90}_{-0.87 }$ &    1.6  &    0.2 &  $.. $ &  $12.85 ^{ 0.18}_{ 0.18 }$  &  $ 3.55  \pm 0.03    $  &$..$ &$..$ &    $..$  \\
18543443+015304 &     MZM72 &        M1 &   2.03 &  $12.58 ^{ 1.06}_{-1.10 }$ &    1.5  &    0.9 &  $.. $ &  $.. $  &  $ 3.48  \pm 0.03    $  &$..$ &$..$ &    $..$  \\
18565849+013452 &     MZM74 &        M1 &   0.96 &  $12.64 ^{ 0.79}_{-0.66 }$ &    1.1  &    1.2 &  $13.25 ^{ 0.19}_{-0.19 }$ &  $13.94 ^{ 0.20}_{ 0.20 }$  &  $ 3.99  \pm 0.03    $  &$..$ &$..$ &    $..$  \\
18580434+021541 &     MZM75 &      M0.5 &   0.91 &  $13.09 ^{ 0.68}_{-0.82 }$ &    1.2  &    1.2 &  $13.49 ^{ 0.22}_{-0.20 }$ &  $14.05 ^{ 0.14}_{ 0.14 }$  &  $ 3.61  \pm 0.03    $  &$..$ & 0.17  &   183.80   \\
19001228+031225 &     MZM77 &        M1 &   0.80 &  $11.72 ^{ 0.52}_{-0.43 }$ &    1.3  &    2.5 &  $13.00 ^{ 0.21}_{-0.22 }$ &  $.. $  &  $ 3.89  \pm 0.02    $  &$..$ & 0.10  &   189.12   \\
19001812+032541 &     MZM78 &      M0.5 &   0.97 &  $12.92 ^{ 0.71}_{-0.67 }$ &    2.0  &    0.4 &  $.. $ &  $13.80 ^{ 0.25}_{ 0.25 }$  &  $ 3.77  \pm 0.03    $  &$..$ &$..$ &    $..$  \\
19060934+055844 &     MZM83 &      K5.5 &   0.74 &  $13.36 ^{ 0.86}_{-0.97 }$ &    2.8  &    0.5 &  $.. $ &  $13.53 ^{ 0.14}_{ 0.14 }$  &  $ 3.93  \pm 0.02    $  & 0.06  & 0.48  &   148.50   \\
19102566+081852 &     MZM84 &        M1 &   0.94 &  $13.07 ^{ 0.78}_{-0.76 }$ &    1.8  &    0.1 &  $.. $ &  $13.33 ^{ 0.16}_{ 0.16 }$  &  $ 3.95  \pm 0.02    $  & 0.09  & 0.64  &   146.01   \\
19125995+094801 &     MZM85 &        M7 &   0.61 &  $13.27 ^{ 1.10}_{-0.74 }$ &    1.6  &   -0.0 &  $.. $ &  $14.07 ^{ 0.18}_{ 0.18 }$  &  $ 3.92  \pm 0.03    $  & 0.39  &$..$ &    $..$  \\
19130113+100159 &     MZM86 &        M1 &   0.87 &  $12.97 ^{ 0.75}_{-0.67 }$ &    2.7  &    0.4 &  $13.89 ^{ 0.47}_{-0.60 }$ &  $13.85 ^{ 0.15}_{ 0.15 }$  &  $ 3.88  \pm 0.03    $  &$..$ & 0.18  &   457.63   \\
19141414+102802 &     MZM87 &      M2.5 &   1.06 &  $12.15 ^{ 0.69}_{-0.57 }$ &    2.2  &    1.5 &  $.. $ &  $14.63 ^{ 0.18}_{ 0.18 }$  &  $ 3.83  \pm 0.02    $  &$..$ &$..$ &    $..$  \\
19214456+133722 &     MZM91 &      M1.5 &   1.03 &  $12.89 ^{ 0.78}_{-0.76 }$ &    1.3  &    1.0 &  $.. $ &  $14.79 ^{ 0.22}_{ 0.22 }$  &  $ 3.96  \pm 0.03    $  & 0.07  &$..$ &    $..$  \\

\hline
\end{tabular} 
}
\end{centering}
\end{myresizeenv}

\begin{list}{}
\item 
2MASS-ID:  Source designation in the 2MASS catalog \citep{cutri03}.  \\
MZM-ID: Source designation in the catalog of \citet{messineo16}. \\
RSG-Sp: Spectral-type from \citet{messineo16}.\\
\Aks:  extinction in \Ks\ band calculated as in \citet{messineo17} and \citet{messineo19}. \\
DM1: distance moduli from \citet[][]{messineo17} (based on clump stars).\\
DM2:  Gaia EDR3 parallactic distances from \citet{bailer21}, as in  \citet{messineo21z}.\\
RUWE: Gaia EDR3 renormalized unit weight error \citep{gaia21}. \\ 
frac: Gaia EDR3 ratio of the parallax values and their errors \citep{gaia21}.\\   
\Ks$_o$: dereddened 2MASS \Ks\ magnitudes, 
as in \citet{messineo17} and \citet{messineo19}.\\   
Ampl(1): estimated Gaia amplitudes from the Gaia DR2 
photometric uncertainty by \citet{mowlavi21}. Ampl(2), Per(2): magnitude amplitudes and periods from 
the AAVSO International Variable Star database (VSX) \citep{AAVSO}. \\
\end{list}

\end{table*}

\section{Observations}
\label{sec.data}
To monitor the  fluxes of the 57 late-type stars, 
the infrared ANDICAM camera mounted on the 1.3m telescope
of Cerro Tololo in Chile was used.
The telescope  is operated by the SMARTS consortium, and a total 
of 28.7 nights were allocated to this program 
(2016B-0106) by the Telescope Allocation Committees 
(TAC) for the National Optical Astronomy Observatory.
A CCD camera was attached for simultaneous 
optical observations.

\begin{figure*}
\begin{center}
\resizebox{0.48\hsize}{!}{\includegraphics[angle=0]{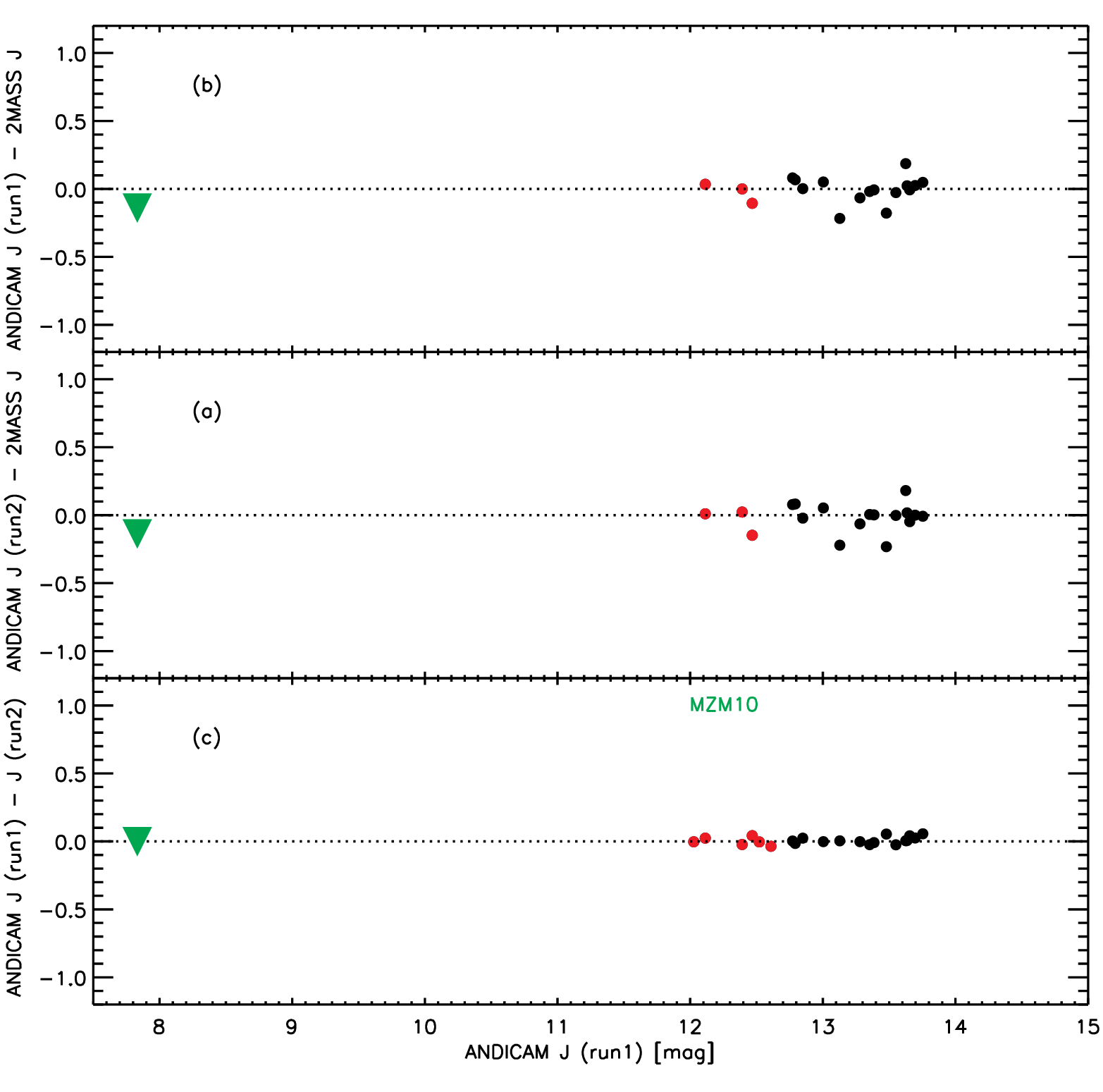}}
\resizebox{0.48\hsize}{!}{\includegraphics[angle=0]{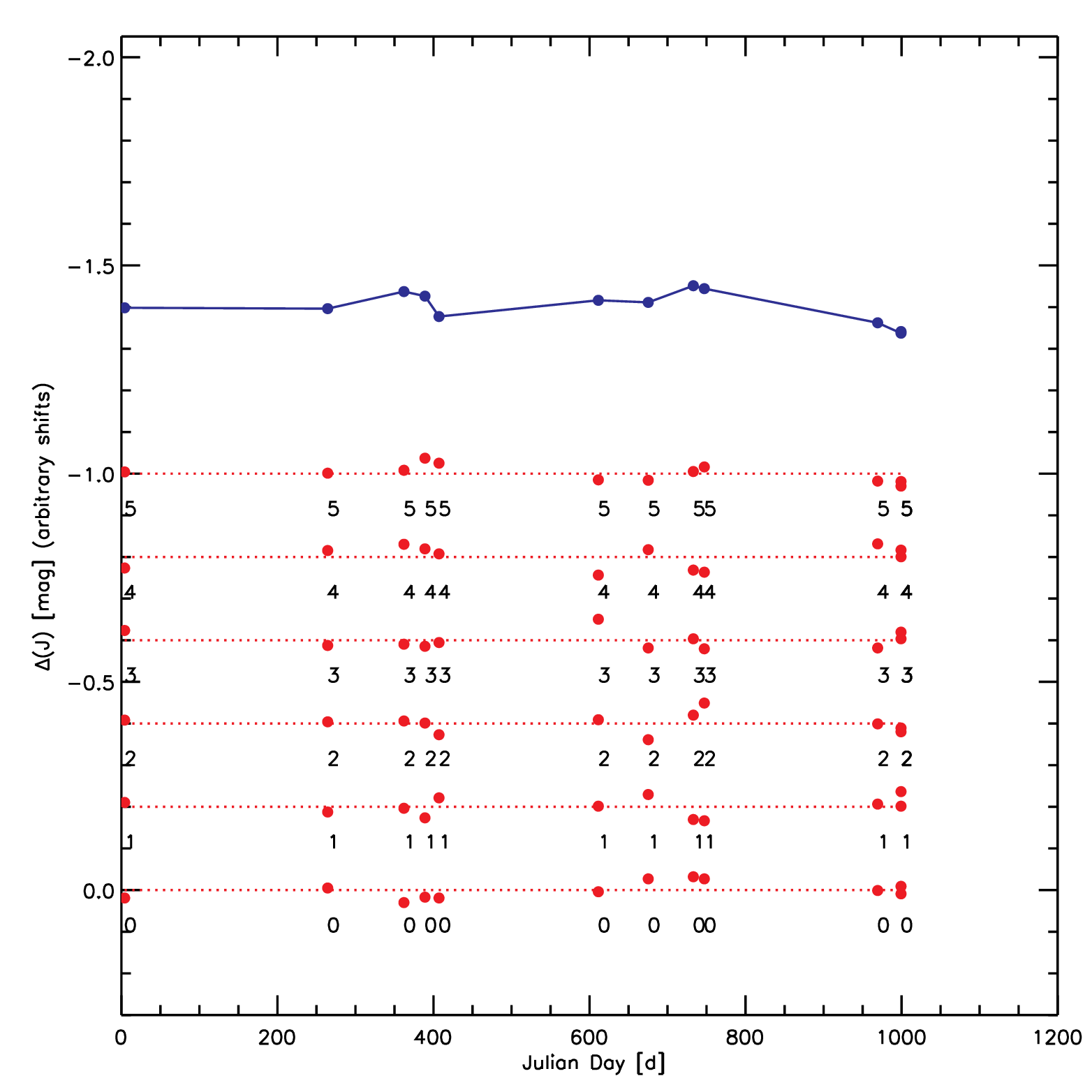}}
\end{center}
\caption{\label{deltaCALJ.fig} \label{delta2massJ.fig} 
{\it Left panel:} Diagram of the ANDICAM $J$ magnitudes 
vs.  the differences between the 2MASS and the ANDICAM $J$
magnitudes (a and b) and between two ANDICAM epochs (c),
in the  field of MZM10. Red-filled circles mark stars used
 for the  photometric calibration.
The green triangle shows the location of MZM10.
{\it Left panel:} 
Magnitude variations, $\Delta$ (J), in $J$-band vs. time 
of the MZM10 star (blue). The $I$-band variations of its photometric 
calibrators are also shown (red). The calibrators are stars taken from 
the same field of view (see text) 
and are marked with 0, 1, 2, 3, 4, and 5.
} 
\end{figure*}

\section{ $J$-band observations}
\label{sec.tables}

The  infrared array of ANDICAM consists of
$1024 \times 1024$ pixels 
($512 \times 512$ pixels with a pixel scale of 
0\farcs276 pix$^{-1}$ after 
a default $2 \times2$ binning)  and has a field of view
of $2\farcm 4 \times 2\farcm 4$.
We used the $J$-filter.
The observing sequence is made with the 
classical 7 dithered positions 
(the dither scale parameter was set to 40,
i.e., each dithered position is within 20\arcsec\
from the center, or, equivalently, the center moves within a box of $40\arcsec \times 40\arcsec$). 
Eight exposures were taken for each star. 
Each exposure consists of two coadds  and their 
integration times ranged from 4-30 s. 
This allowed us to obtain an excellent 
sky subtraction (using a robust mean with a 2 $\sigma$ clipping).
Each frame was sky-subtracted and flat-fielded.

The world coordinate system of each $J$-band exposure
was created using, as a  reference system,
the coordinates of bright 2MASS $J$ point sources
($J >14$ mag) detected by ANDICAM.

The peaks of the stellar counts vary from a few hundred 
to 3000, as recommended in the ANDICAM manual. 
A few frames were discarded because of saturation.

Typically, the full width half-maximum (FWHM) of the PSF ranged  from 1\arcsec\ to 1\farcs4. 
Aperture photometry was performed using the DAOPHOT \citep{stetson87} 
version available in the NASA IDL Astronomy User's Library 
\citep{landsman93}. 
Aperture photometry was performed on each frame 
(at each dithered position) as the targeted stars are among the
brightest in the field and the derived magnitudes were averaged. 

\subsection{$J$-band flux calibration}
Photometric calibration was performed in a relative manner. 
The measured magnitudes were registered on those
measured in the first (reference) epoch.
The absolute calibration was done using field stars 
with known 2MASS $J$-band magnitudes brighter than 13 mag. 
The calibrators were visually inspected and a few stars with 
a larger dispersion discarded
(see Fig.\ \ref{delta2massJ.fig}).
%The DENIS $J$-band magnitudes are usually consistent with those from 2MASS 
%\citep[with a mean difference of 0.01 and  a standard deviation of 0.12 mag,][]{schultheis01,messineo04}.
%The photometric calibration was repeated after having selected 
%isolated non-variable bright field stars, as shown in Fig. \ref{deltaCALJ.fig}. 
In a given field, the average  standard deviation of the calibrator ANDICAM $J$ magnitudes 
ranges from 0.001-0.044 mag, as listed { in} Table \ref{table.mag}.\\
The average $J$-band  magnitudes of the targets 
range from 6.9 to 11.6 mag, as listed in Table \ref{table.mag}.

\begin{sidewaystable*}
\caption{\label{table.mag} The average $J$ and $I$ magnitudes of the observed stars. }
\renewcommand{\arraystretch}{0.6}
\begin{myresizeenv}
\begin{centering}
{\tiny
%\begin{tabular}{@{\extracolsep{-.08in}}l|rrrrrr|rrrrrrrrrrrrrr|}
\begin{tabular}{l|rrrrrr|rrrrrrrrrrrrrr|}
\hline
\hline
     &        \multicolumn{6}{c}{\rm ANDICAM $J$} & \multicolumn{7}{c}{\rm ANDICAM $I$}\\
NAME & Nobs(J)&Ncal(J)&$<$std\_cal(J)$>$&$<$mag$_*$(J)$>$&$\sigma_*(J)$& $\sigma_{ext}$&
Nobs(I)&Ncal(I)&$<$std\_cal(I)$>$&$<$mag$_*$(I)$>$&$\sigma_*(I)$&
 $<\sigma I_{circa}>$& $\sigma_{ext}$ & $\Delta_{DEN-PAN}$ & $\Delta_{DEN-VPHAS}$\\
     &        &       &            &                &           &           & 
       &       &           &                &         &           &    &  C$_{\rm A}$          \\ 
     &        &       & [mag]           &  [mag]              & [mag]    &[mag]                
       &       &                           & [mag]          & [mag]          & [mag]   & [mag] &[mag]  &[mag] &[mag]          \\ 
\hline
18112728-175014  &   11 &   3 &   0.020 &   8.420 &   0.027 &   0.087 &   13 &   5 &   0.018  &  15.687  &   0.133  &   0.103  &   0.038  &  -0.084  &   0.003 \\
18114736-192915  &   19 &   2 &   0.017 &   9.351 &   0.016 &   0.002 &     &    &       &       &       &       &       &       &      \\
18131562-180122  &   13 &   4 &   0.015 &   8.017 &   0.039 &   0.027 &   10 &   4 &   0.009  &  15.311  &   0.126  &   0.050  &   0.037  &  -0.116  &  -0.063 \\
18132341-185818  &   12 &   8 &   0.024 &   8.851 &   0.038 &   0.057 &   10 &   5 &   0.025  &  15.054  &   0.051  &   0.045  &   0.023  &  -0.109  &  -0.046 \\
18134914-184633  &   13 &   2 &   0.012 &   8.280 &   0.035 &   0.013 &   13 &   7 &   0.026  &  15.480  &   0.157  &   0.071  &   0.092  &  -0.157  &  -0.178 \\
18140682-190620  &   12 &   6 &   0.024 &   7.830 &   0.039 &   0.073 &   12 &   8 &   0.028  &  15.043  &   0.118  &   0.053  &   0.064  &  -0.119  &  -0.046 \\
18144593-173754  &   13 &   6 &   0.015 &   9.809 &   0.042 &   0.035 &     &    &       &       &       &       &       &       &      \\
18154112-164645  &   13 &   5 &   0.026 &   7.979 &   0.044 &   0.019 &   11 &   6 &   0.023  &  14.497  &   0.097  &   0.042  &   0.058  &  -0.151  &  -0.123 \\
18155832-165827  &   34 &   5 &   0.028 &   9.219 &   0.034 &   0.022 &     &    &       &       &       &       &       &       &      \\
18171598-140554  &   12 &   4 &   0.024 &   7.340 &   0.021 &   0.039 &   12 &   5 &   0.017  &  13.232  &   0.052  &   0.021  &   0.073  &  -0.060  &  -0.084 \\
18172865-163739  &   14 &   2 &   0.021 &   8.516 &   0.026 &   0.002 &   10 &   7 &   0.021  &  15.927  &   0.096  &   0.090  &   0.080  &  -0.122  &  -0.081 \\
18174160-135628  &    7 &   3 &   0.026 &   7.650 &   0.029 &   0.044 &   11 &   6 &   0.031  &  14.187  &   0.119  &   0.037  &   0.106  &  -0.078  &  -0.060 \\
18175212-171508  &   13 &   5 &   0.022 &  10.279 &   0.021 &   0.031 &     &    &       &       &       &       &       &       &      \\
18184453-165108  &   12 &   1 &   0.001 &  11.305 &   0.019 &   0.000 &     &    &       &       &       &       &       &       &      \\
18184660-163456  &   14 &   1 &   0.020 &   9.089 &   0.027 &   0.000 &    3 &  10 &   0.012  &  17.488  &   0.149  &   0.077  &   0.084  &  -0.048  &  -0.054 \\
18210685-150340  &   14 &   1 &   0.007 &   8.126 &   0.045 &   0.000 &   13 &   9 &   0.015  &  15.322  &   0.157  &   0.135  &   0.044  &  -0.151  &  -0.152 \\
18210846-153209  &   11 &   4 &   0.018 &   8.358 &   0.037 &   0.048 &   11 &   4 &   0.040  &  15.087  &   0.147  &   0.079  &   0.103  &  -0.210  &  -0.204 \\
18212427-135528  &   12 &   9 &   0.030 &   7.515 &   0.027 &   0.032 &   12 &   6 &   0.026  &  13.587  &   0.101  &   0.082  &   0.047  &  -0.102  &  -0.081 \\
18230411-135416  &    9 &   5 &   0.030 &   7.671 &   0.022 &   0.053 &   10 &   4 &   0.020  &  13.757  &   0.032  &   0.033  &   0.150  &  -0.121  &  -0.189 \\
18231119-135758  &   13 &   2 &   0.030 &   7.826 &   0.033 &   0.029 &   11 &   4 &   0.021  &  14.337  &   0.060  &   0.084  &   0.059  &  -0.117  &  -0.157 \\
18240991-110121  &    9 &   8 &   0.023 &   6.903 &   0.022 &   0.050 &   10 &   6 &   0.023  &  11.533  &   0.056  &   0.042  &   0.119  &  -0.152  &  -0.157 \\
18251120-114056  &   10 &   4 &   0.032 &   7.036 &   0.043 &   0.065 &   12 &   6 &   0.017  &  12.690  &   0.209  &   0.022  &   0.097  &  -0.209  &  -0.188 \\
18254382-115336  &    9 &  11 &   0.027 &   6.853 &   0.012 &   0.095 &   11 &   9 &   0.021  &  10.818  &   0.015  &   0.022  &   0.062  &  -0.224  &  -0.190 \\
18261922-140648  &   12 &   3 &   0.027 &   7.485 &   0.099 &   0.113 &   12 &  27 &   0.023  &  13.689  &   0.158  &   0.032  &   0.043  &  -0.198  &  -0.196 \\
18291233-121940  &   12 &   4 &   0.017 &   7.523 &   0.010 &   0.086 &   10 &   6 &   0.029  &  12.639  &   0.018  &   0.024  &   0.069  &  -0.160  &  -0.181 \\
18301889-102000  &   11 &   6 &   0.020 &   6.975 &   0.026 &   0.073 &   10 &   5 &   0.028  &  11.865  &   0.087  &   0.045  &   0.116  &  -0.302  &  -0.256 \\
18310460-105426  &   13 &   8 &   0.029 &   7.859 &   0.039 &   0.067 &   12 &   7 &   0.030  &  14.658  &   0.062  &   0.042  &   0.058  &  -0.224  &  -0.237 \\
18315881-111921  &   12 &   2 &   0.024 &   7.308 &   0.036 &   0.028 &   12 &   8 &   0.012  &  13.561  &   0.063  &   0.023  &   0.027  &  -0.192  &  -0.160 \\
18334444-065947  &   11 &   5 &   0.033 &   7.169 &   0.065 &   0.062 &   12 &   4 &   0.015  &  13.174  &   0.227  &   0.030  &   0.030  &  -0.166  &  -0.147 \\
18352902-072112  &   13 &   2 &   0.006 &   9.929 &   0.030 &   0.011 &     &    &       &       &       &       &       &       &      \\
18353475-075648  &   55 &   5 &   0.025 &   9.118 &   0.031 &   0.038 &     &    &       &       &       &       &       &       &      \\
18354911-073443  &   13 &   6 &   0.024 &   8.686 &   0.067 &   0.034 &    8 &   7 &   0.016  &  16.157  &   0.084  &   0.089  &   0.058  &  -0.167  &  -0.160 \\
18355151-073011  &   14 &   1 &   0.071 &  10.083 &   0.030 &   0.000 &     &    &       &       &       &       &       &       &      \\
18374651-071224  &   13 &   2 &   0.018 &   9.743 &   0.023 &   0.004 &     &    &       &       &       &       &       &       &      \\
18413481-044857  &   13 &   4 &   0.035 &   7.849 &   0.037 &   0.148 &   11 &  10 &   0.027  &  13.848  &   0.079  &   0.029  &   0.069  &  -0.254  &  -0.279 \\
18414834-044852  &   12 &   2 &   0.024 &   8.180 &   0.023 &   0.010 &   11 &  11 &   0.028  &  14.571  &   0.063  &   0.034  &   0.076  &  -0.263  &  -0.316 \\
18421710-044116  &   13 &   3 &   0.024 &  11.426 &   0.064 &   0.029 &     &    &       &       &       &       &       &       &      \\
18424231-044053  &   13 &   5 &   0.024 &   7.519 &   0.031 &   0.021 &   12 &   8 &   0.015  &  13.181  &   0.085  &   0.125  &   0.062  &  -0.171  &  -0.216 \\
18424479-043357  &   12 &   4 &   0.034 &   7.105 &   0.030 &   0.059 &   11 &  13 &   0.021  &  11.983  &   0.039  &   0.024  &   0.050  &  -0.237  &  -0.239 \\
18425222-034618  &   14 &   5 &   0.037 &   9.250 &   0.036 &   0.027 &     &    &       &       &       &       &       &       &      \\
18430800-035624  &   12 &   5 &   0.018 &  11.650 &   0.027 &   0.019 &     &    &       &       &       &       &       &       &      \\
18442616-033527  &   12 &   2 &   0.044 &   8.504 &   0.067 &   0.004 &    9 &   9 &   0.030  &  15.959  &   0.101  &   0.073  &   0.073  &  -0.283  &  -0.280 \\
18464441-032404  &   15 &   4 &   0.027 &   9.321 &   0.027 &   0.071 &     &    &       &       &       &       &       &       &      \\
18464480-030332  &   10 &   2 &   0.026 &   7.875 &   0.031 &   0.007 &   11 &   2 &   0.014  &  14.333  &   0.032  &   0.033  &   0.115  &  -0.339  &  -0.293 \\
18475357-014715  &   13 &   9 &   0.020 &   8.890 &   0.015 &   0.046 &    4 &   7 &   0.009  &  16.472  &   0.091  &   0.095  &   0.067  &  -0.379  &  -0.353 \\
18482997-021150  &   14 &   2 &   0.044 &   8.039 &   0.036 &   0.034 &     &    &       &       &       &       &       &       &      \\
18543443+015304  &   14 &   1 &   0.000 &  10.299 &   0.046 &   0.000 &     &    &       &       &       &       &       &       &      \\
18565849+013452  &    9 &   8 &   0.030 &   7.667 &   0.017 &   0.593 &   12 &   4 &   0.016  &  12.887  &   0.025  &   0.020  &   0.018  &  -0.172  &  -0.183 \\
18580434+021541  &   12 &   3 &   0.038 &   7.248 &   0.050 &   0.092 &   11 &  10 &   0.027  &  12.471  &   0.134  &   0.030  &   0.133  &       &      \\
19001228+031225  &   11 &   4 &   0.041 &   7.030 &   0.020 &   0.149 &   10 &  10 &   0.035  &  11.904  &   0.010  &   3.450  &   0.075  &       &      \\
19001812+032541  &   10 &   2 &   0.024 &   7.467 &   0.040 &   0.011 &    6 &   5 &   0.025  &  13.981  &   0.083  &   0.048  &   0.222  &       &      \\
19060934+055844  &   10 &   1 &   0.022 &   7.024 &   0.067 &   0.000 &   11 &   3 &   0.037  &  12.374  &   0.163  &   0.116  &   0.090  &       &      \\
19102566+081852  &    9 &   2 &   0.017 &   7.528 &   0.049 &   0.007 &   11 &   3 &   0.029  &  13.504  &   0.138  &   0.028  &   0.305  &       &      \\
19125995+094801  &   11 &   6 &   0.035 &   7.170 &   0.172 &   0.044 &   10 &  19 &   0.025  &  12.623  &   0.366  &   0.037  &   0.170  &       &      \\
19130113+100159  &   12 &   4 &   0.038 &   7.302 &   0.031 &   0.065 &   12 &  20 &   0.031  &  12.513  &   0.046  &   0.039  &   0.116  &       &      \\
19141414+102802  &   14 &   2 &   0.017 &   7.907 &   0.020 &   0.007 &    9 &  18 &   0.027  &  14.202  &   0.031  &   0.040  &   0.480  &       &      \\
19214456+133722  &   12 &   2 &   0.042 &   7.874 &   0.036 &   0.008 &    9 &  12 &   0.023  &  13.720  &   0.053  &   0.032  &   0.142  &       &      \\

\hline
\end{tabular}
}
\end{centering}
\end{myresizeenv}
\begin{list}{}
\item {\bf Notes:} Nobs(J) = number of used $J$-band observations; 
Ncal(J) = number of surrounding stars used for the photometric calibration;
$<$std\_cal(J)$>$ = mean standard deviation of the calibrator magnitudes;
%$\sigma$(std\_cal(J)) = $\sigma$ of the standard deviations 
%of the calibrator $J$ magnitudes; 
$<$mag$_*$(J)$>$ = $J$-band average magnitude of the target;
$\sigma_*(J)$ = standard deviation of the target magnitudes;
$\sigma_{ext}$ = the standard deviation of 
the differences between the ANDICAM $J$ magnitudes 
(reference epoch) and  the 2MASS $J$ magnitudes  
of the calibrator stars;
Nobs(I) = number of used $J$-band observations; 
Ncal(I) = number of surrounding stars used for the photometric calibration;
$<$std\_cal(I)$>$ = mean standard deviation of the calibrator magnitudes;
%$\sigma$(std\_cal(I)) = $\sigma$ of the standard deviations of 
%the calibrator magnitudes;
$<$mag$_*$(I)$>$ = $I$-band average magnitude of the target;
$\sigma_*(I)$ = standard deviation of the target magnitudes; 
$<\sigma I_{circa}>$ = mean of the standard deviations of field 
stars with magnitudes similar to that of the targeted late-type star;
%$\sigma(\sigma I_{circa})$ = standard deviation of $\sigma I_{circa}$ 
%of field stars with magnitudes similar to that of the target.
$\sigma_{ext}$ = the standard deviation of 
 the differences between the ANDICAM $I$ magnitudes 
(reference epoch) and  the $I$ magnitudes from 
the external catalogs (DENIS or Pan-STARRS).
$\Delta_{DEN-PAN}$=  median of $($DENIS $I -$ Pan-STARRS $I_{\rm AB}$ + 0.445$)$
of field stars with $I<$ 17.0 mag.
$\Delta_{DEN-VPHAS}$=  median of $($DENIS $I -$ VPHAS $I_{\rm Vega}$  $)$
of field stars with $I<$ 17.0 mag.

\end{list}
\end{sidewaystable*}

\renewcommand{\arraystretch}{0.9}
\begin{table*}
\caption{\label{mag-compare-target}  Magnitude variations in $I$- and $J$-bands of the targeted stars.}
{\tiny
\begin{tabular}{@{\extracolsep{-.08in}}lr|rr|rr|rrrrrrr}
 \hline
 \hline
     2MASS-ID       & MZM &  ANDICAM             & ANDICAM $<$J$>$ & ANDICAM              & ANDICAM $<$I$>$ & fl\_corr & fl\_var & Period & fap & $\Delta$I/2 \\
                    &     & J(max)$-$J(min)      &  $-$ 2MASS J      &I(max) $-$ I(min)   &$-$ DENIS I      &  &\\
                    &     &   [mag]              &    [mag]          &[mag]               &[mag]            &          &         & [d]    &     & [mag]\\                                                                       &         &        &[d]  & [\%]& [mag]  \\
\hline

    18131562-180122 &          MZM06 &   0.127 &  $-$0.157 &   0.378 &   0.284 &  1  & 1  & 248   & 23   &   0.191  &  \\
    18134914-184633 &          MZM09 &   0.123 &  $-$0.103 &   0.453 &   0.567 &  1  & 1  & 433   & 11   &   0.167  &  \\
    18210685-150340 &          MZM33 &   0.182 &  $-$0.260 &   0.567 &   0.164 &  1  & 1  & 201   & 24   &   0.195  &  \\
    18210846-153209 &          MZM34 &   0.116 &  $-$0.045 &   0.487 &   0.151 &  1  & 1  & 167   &  6   &   0.184  &  \\
    18251120-114056 &          MZM40 &   0.109 &  $-$0.025 &   0.578 &  $-$0.065 &  1  & 1  & 432   & 13   &   0.215  &  \\
    18334444-065947 &          MZM50 &   0.190 &  $-$0.069 &   0.692 &   0.246 &  1  & 1  & 431   & 23   &   0.218  &  \\
    19060934+055844 &          MZM83 &   0.236 &  $-$0.006 &   0.530 &    &  1  & 1  & 279   & 43   &   0.166  &  \\
    19125995+094801 &          MZM85 &   0.517 &   0.067 &   1.076 &    &  1  & 1  & 260   & 23   &   0.453  &  \\

\hline
\end{tabular}
\begin{list}{}
\item ANDICAM $J$(max)$-J$(min)  is the difference between the
maximum and minimum ANDICAM $J$ magnitudes.\\
ANDICAM $<J>$ $-$ 2MASS $J$ is the difference between the average
ANDICAM $J$  and the 2MASS $J$ magnitudes.\\
ANDICAM $I$(max)$-I$(min)  is the difference between the
maximum and minimum ANDICAM $I$ magnitudes.\\
ANDICAM $<I>$ $-$ DENIS $I$ is the difference between the average
ANDICAM $I$  and the DENIS $I$ magnitudes.\\
fl\_corr =1 if the Pearson correlation 
coefficient of the simultaneously taken $I$ mag and $J$ mag vectors 
is larger than 0.5.\\
fl\_var =1 if the standard deviation of the $I$ mag vector
exceeds  twice  that of field stars at similar magnitudes or 
if the standard deviation of the $J$ mag vector exceeds 2 times
that of other bright field stars (non-variable calibrators).\\
Period of the periodic light variation detected in 
the Lomb-Scargle peridiogram (see text).\\
fap is the false alarm probability corresponding to the power level
of the adopted period. \\
$\Delta$I/2 is the semi-amplitude of the photometric $I$-band 
light curve. \\
\end{list}
}
\end{table*} 

\section{ $I$-band observations}
The ANDICAM  CCD detector was also used for simultaneous 
$I$-band observations, taken in staring mode.
The detector with $1024 \times 1024$ pixels 
covers a field of view of
$6\farcm 33 \times 6\farcm 33$. 
For each observation,  from 6 to 7 frames  were acquired. 
The integration time was set to 8 s, and we reached  
a peak of 50 counts on a 16 mag star with
a seeing of 1\farcs1.\\
The CCD frames are distributed by the observatory
after corrections for bias and flat field.
The individual frames were combined 
with a 10 $\sigma$ clipping.

For the astrometric calibration of each observation, 
DENIS data points brighter
than $I$ = 15 mag (or 13 mag in the denser field)
were overlaid on the CCD image.
%From 7 up to 38 DENIS $I$-band data points were 
%visually paired with ANDICAM  data points,
%to create  a   world coordinate system (wcs).
The absolute astrometric solution is $\approx$0\farcs15 accurate.
For the 12 missing fields, 2MASS $J$-band data points were used.

Aperture photometry was performed using the daophot \citep{stetson87} 
version available in the NASA IDL Astronomy User's Library (astron). 
For each field, the 
FWHM was measured and the  aperture radius set to FWHM*0.5+1.5 pix,
and the sky annulus taken from FWHM*0.5+2 to FWHM*0.5+4 pix.

\subsection{$I$-band flux calibration}

The target magnitude calibration was done in a relative manner 
by using field stars with known flux;
the absolute calibration only affects the global zero point,
but not the magnitude variations. 

\begin{figure*}
\begin{center}
\resizebox{0.48\hsize}{!}{\includegraphics[angle=0]{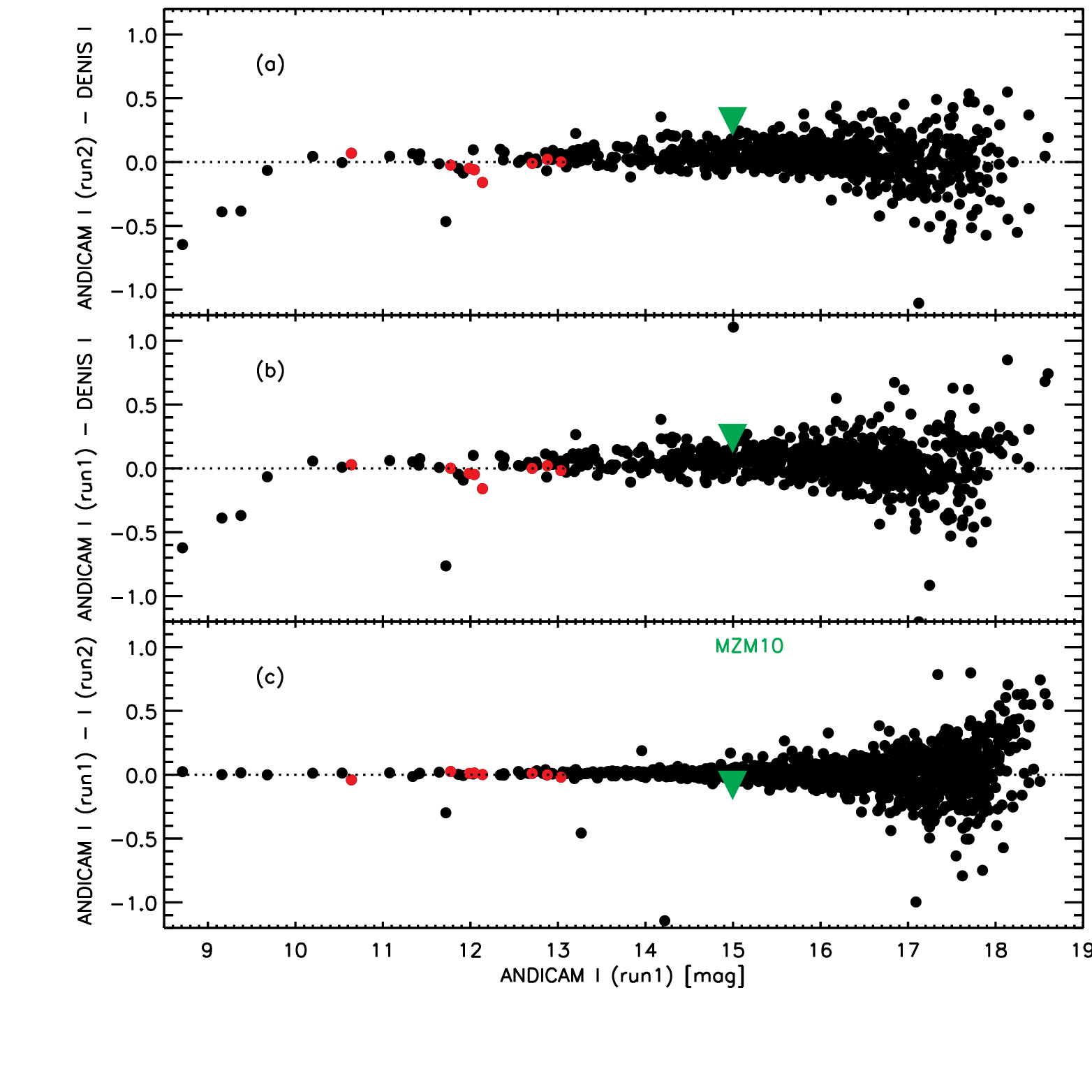}}
\resizebox{0.48\hsize}{!}{\includegraphics[angle=0]{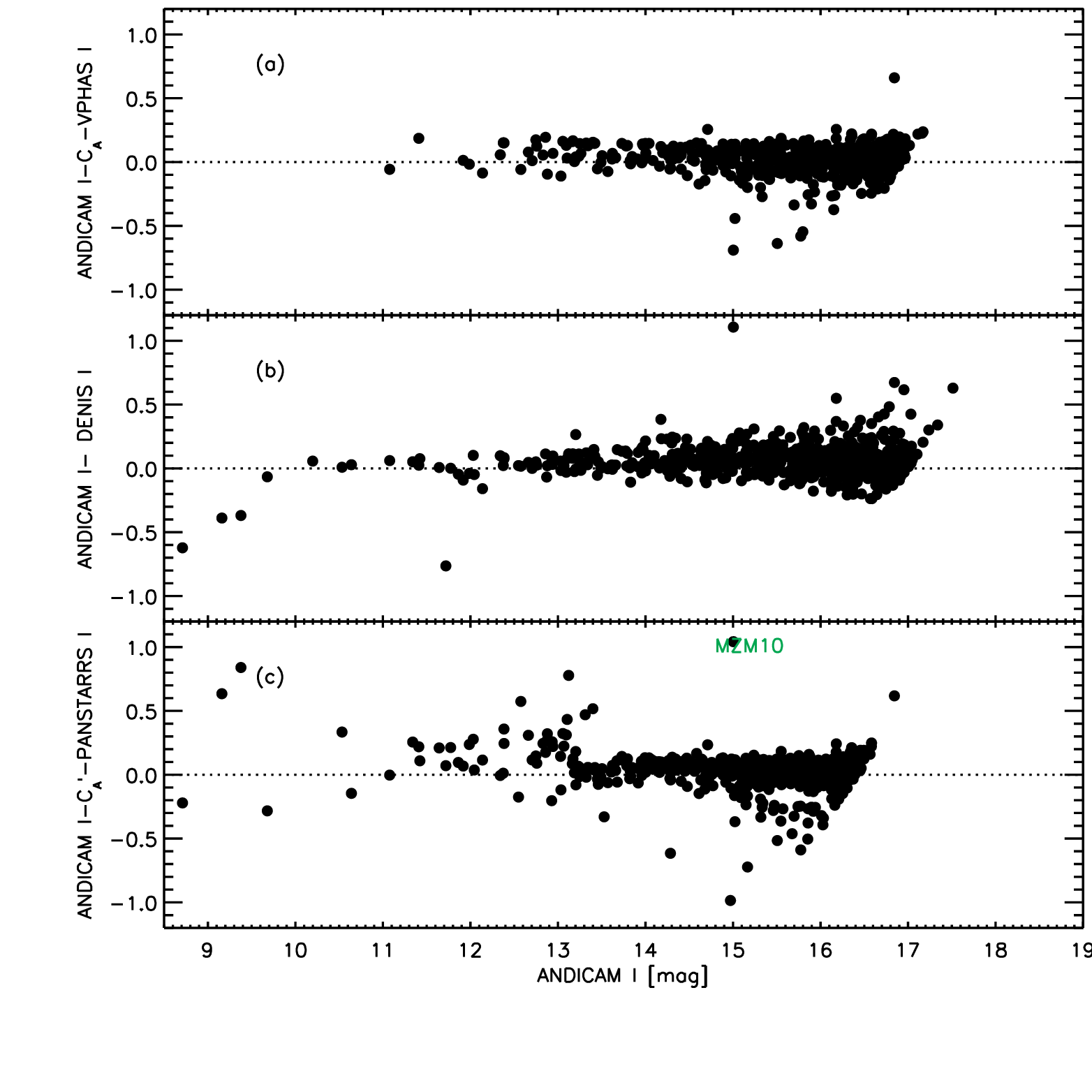}}
\end{center}
\caption{\label{deltaDEN.fig} {\it Right panel:} 
The differences between the DENIS and the ANDICAM $I$
magnitudes (a and b) and between two ANDICAM epochs (c)
are plotted vs.  the ANDICAM $I$ magnitudes. 
Red-filled circles mark stars used for the  photometric calibration.
The green triangle shows the location of MZM21.
{\it Left panel:} For the first epoch of field MZM21, ANDICAM magnitudes  
are compared with VPHAS (a), DENIS (b), and Pan-STARRS (c)  magnitudes.
The constant C$_{\rm A}$ values are as specified in 
Table  \ref{table.mag} and and C$_{\rm A'}$= C$_{\rm A}-0.445$.} 
\end{figure*}

For every epoch, the extracted catalog of point source 
was cross-correlated with 
that of the reference epoch (usually the first epoch)
and flux calibrated.
Most of the observed fields (57 minus 12) 
were covered by the DENIS survey \citep{epchtein94}.
The DENIS  observations in the Gunn-I filter 
saturate at around  10 mag 
and have a 3 $\sigma$ detection limit at 19 mag, e.g., 
as described in \citet{messineo04},  
as shown in Fig.\ \ref{deltaDEN.fig}.
Therefore, DENIS fully covers the range of interest, as the 
$I$ magnitudes of the detected targets 
range from 10.8-17.5 mag.
DENIS point sources with $10 < $ $I$-mag $<  13 $ 
mag were used  to determine the night zero point; 
the median of the differences between the 
instrumental magnitudes and the DENIS $I$-mag was adopted.

\begin{figure*}
\begin{center}
\resizebox{0.49\hsize}{!}{\includegraphics[angle=0]{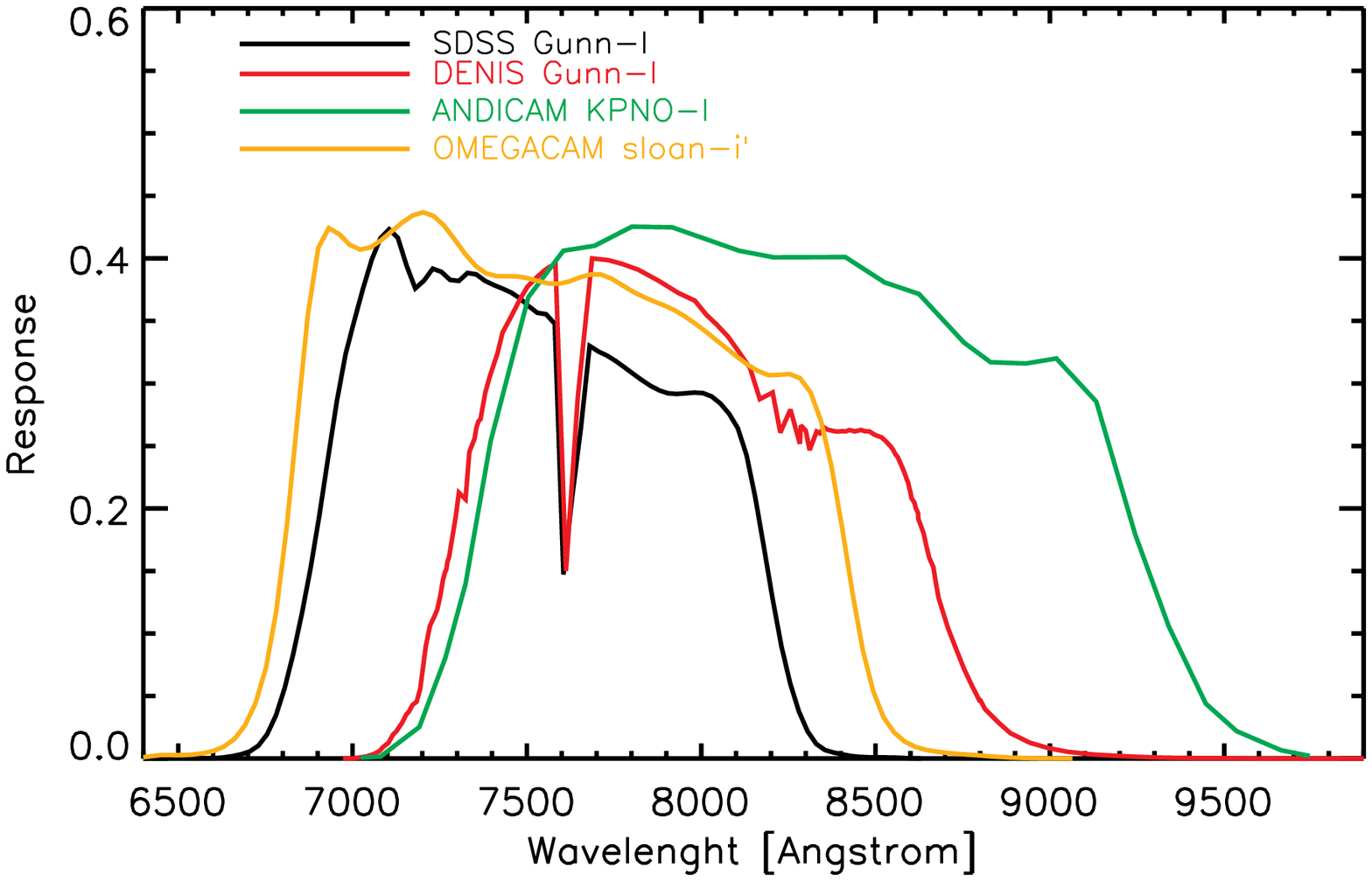}}
\resizebox{0.49\hsize}{!}{\includegraphics[angle=0]{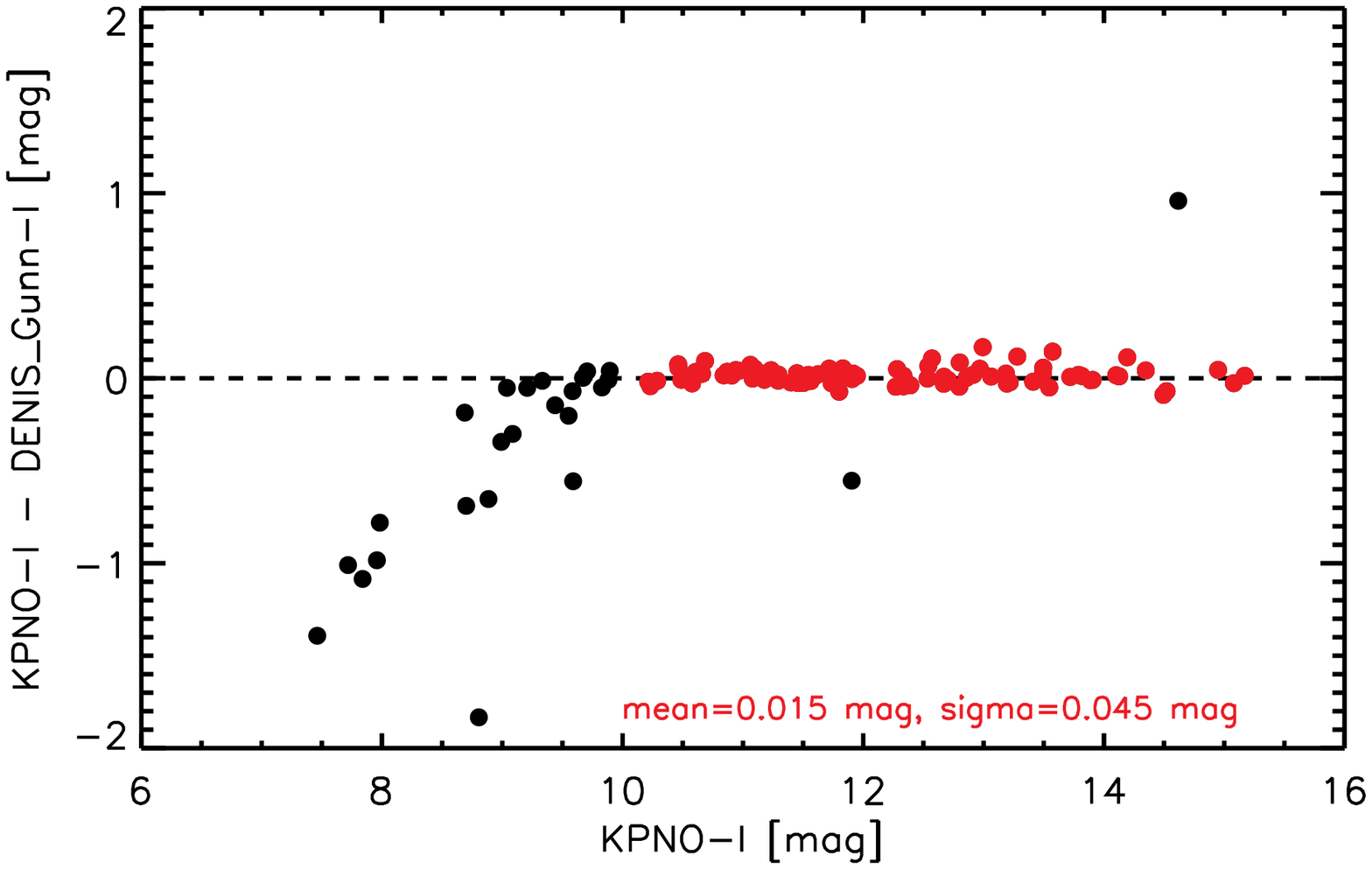}}
\resizebox{0.49\hsize}{!}{\includegraphics[angle=0]{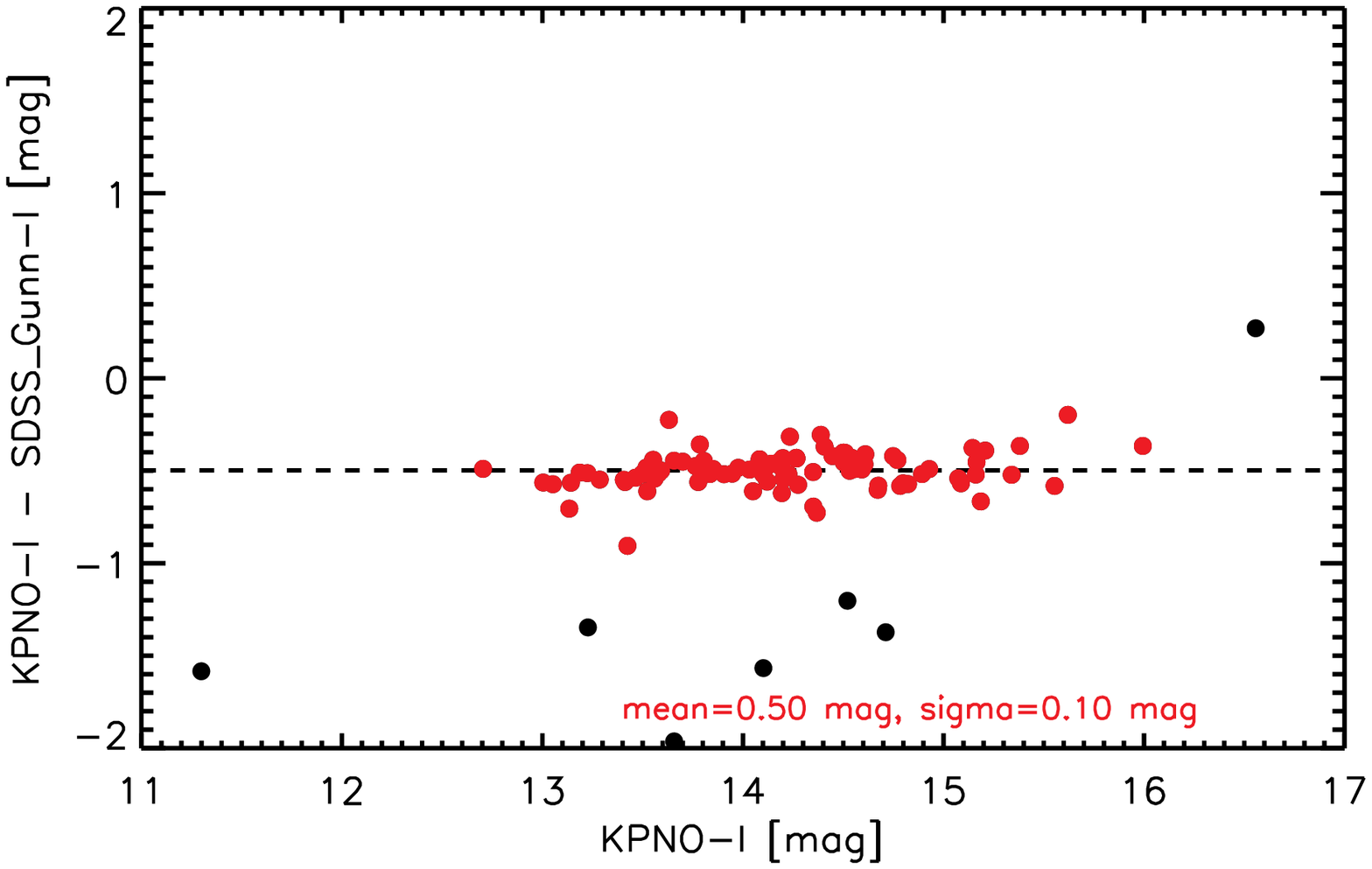}}
\resizebox{0.49\hsize}{!}{\includegraphics[angle=0]{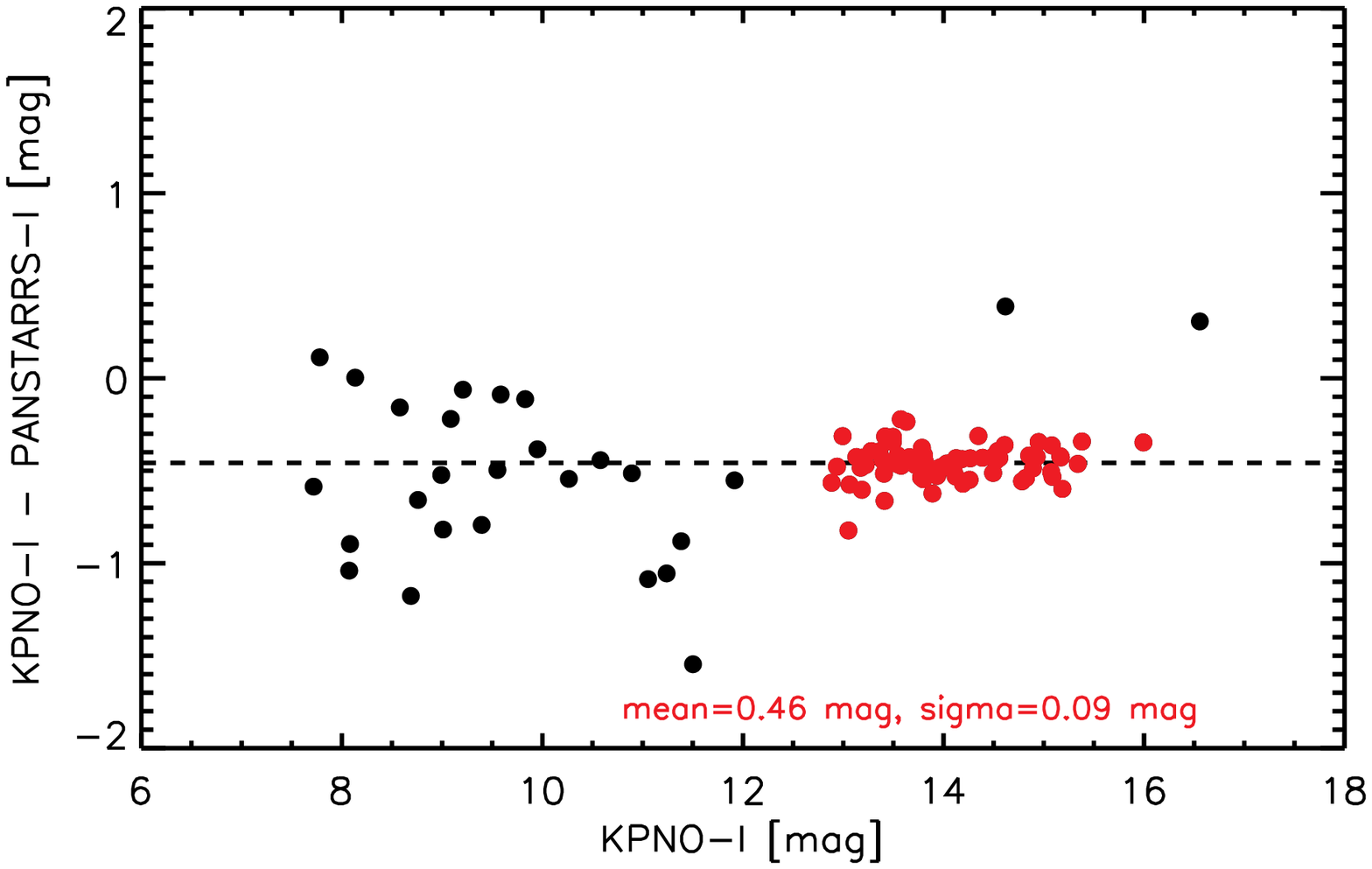}}
\end{center}
\caption{\label{iband_std} The {\it left upper panel} displays the ANDICAM KPNO-I filter 
response in green, that of the DENIS Gunn-I filter in red, 
that of the SDSS Gunn-I filter in black, and that 
of the VPHAS sloan-i' in orange.
In the {\it right upper panel}, the differences between the 
$I$-band magnitudes of the standard stars by \citet{landolt09} 
(in the Johnson-Kron-Cousins system) and the DENIS $I$-band magnitudes are plotted.
In the {\it right lower panel}, there are the differences between the 
$I$-band magnitudes by \citet{landolt09} (in the Johnson-Kron-Cousins 
system) and the SDSS $I$-band magnitudes, and in the {\it left lower panel}, 
 the differences between the $I$-band magnitudes by Landolt's system and 
the Pan-STARRS $I$-band magnitudes.
} 
\end{figure*}

ANDICAM mounts a KPNO-I filter, while the DENIS survey made 
use of a Gunn-I filter, as shown in Fig.\ \ref{iband_std}.
The average difference between the $I$-band 
magnitudes of the standard stars by \citet{landolt09} in the 
Johnson-Kron-Cousins system and the DENIS $I$-band magnitudes
is 0.015 mag with  $\sigma$ = 0.045 mag. 
This offset was not applied.

The average difference between the $I$-band 
magnitudes by \citet{landolt09} and the SDSS $I$-band magnitudes
is 0.50 mag with  $\sigma$ = 0.10 mag when 16$<I$-band $<13$ mag.
However, unfortunately, only three target fields were covered by the 
SDSS survey DR12 \citep{alam15}.
The average difference between the $I$-band 
magnitudes by \citet{landolt09} and the
Pan-STARRS   $I$-band  magnitudes (Mean PSF AB) 
from 16-13 mag is 0.46 with  $\sigma$ = 0.09 mag. 
The standard stars by \citet{landolt09} 
are not covered by the Galactic Plane VPHAS+ survey \citep{drew14}.
However, for most of the observed fields, $I$ magnitudes are 
available from the VPHAS+, and average shifts from 0 to 0.3 mag
are measured between the DENIS and VPHAS+ $I$ (Vega system)
magnitudes.
The bright tail of stars detected in ANDICAM 
were saturated in Pan-STARRS, and VPHAS+ $I$-band catalogs. 

\begin{figure}
\begin{center}
\resizebox{0.98\hsize}{!}{\includegraphics[angle=0]{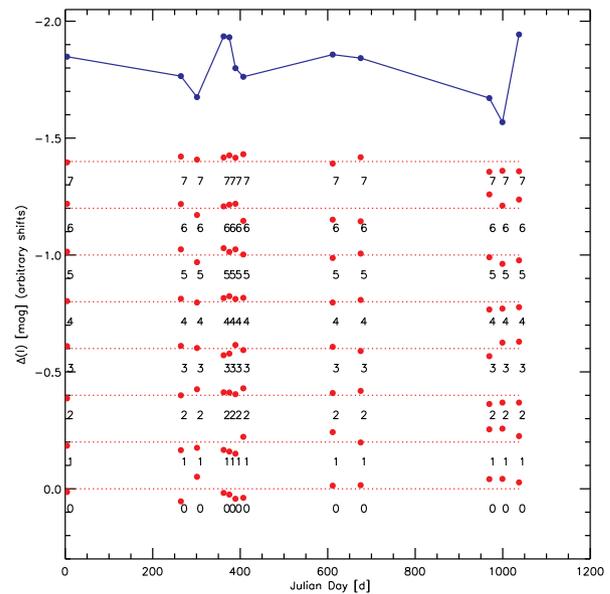}}
\end{center}
\caption{\label{deltaCAL.fig} Magnitude variations, $\Delta$ (I), in $I$-band vs. time 
of the MZM10 star (blue). The $I$-band variations of its photometric 
calibrators are also shown (red). The calibrators are stars taken from 
the same field of view (see text) and are marked with 0, 1, 2, 3, 4, 5, 6, and 7.
} 
\end{figure}

The absolute photometric calibration was  refined
by analyzing the time behaviours of
field  stars    with DENIS $13< I <10$  mag, 
and retaining  as calibrators those  stars 
with smaller ANDICAM time variations (std\_cal(I)$_j$), i.e., 
with std\_cal(I)$_j$ values within the field mean $<$std\_cal(I)$>$ plus
1.5 times their dispersion; std\_cal(I)$_j$ varies from 0.013 mag to 0.048 mag.\\
For 12 fields calibrated with Pan-STARRS (because not covered by DENIS),
stars from $15.5< I <13$  mag were used.
An example of adopted
calibrators is illustrated  in Fig.\ \ref{deltaCAL.fig}.
The computed average $I$-band magnitudes of the targets 
are listed in Table \ref{table.mag}, along with some parameters
(e.g., internal standard deviations of calibrator magnitudes,
and external standard deviations of field stars detected in 
$I$-band by ANDICAM as well as by the DENIS, VPHAS+, and 
Pan-STARRS surveys) to illustrate the uncertainties on the 
absolute calibration.

Only fields where the targeted stars were detected 
in at least 2 epochs were further analyzed (15 stars 
were below the detection threshold).

\section{$J$-band and $I$-band variations}
\label{significant}
In order to assess the existence of significant 
variations in the brightness of the targeted stars,
the $\sigma$ of the target $J$ magnitudes, $\sigma_*(J)$,
are compared with the  $\sigma$ of the calibrator stars;
indeed, the targets are among the brightest stars detected 
in $J$-band. In the  $I$-band, the targets
are  faint; therefore, a $\sigma I_{circa}$ is 
calculated with field stars at the target $I$ magnitude,
and compared with $\sigma_*(I)$.

In the $J$-band analysis, three targets (MZM42, MZM56, and MZM85)
have $\sigma_*(J)$ values larger than 3.5 times the 
$<$std\_cal(J)$>$ of their
calibrators, and larger than their quoted error bars; 
this calculation permits the detections of
variations larger than 0.07 mag, because
the mean of the variations of the calibrators is 0.025 mag.
 In the $I$-band data,
six stars appear significantly variable ($>3.5$ $<\sigma I_{circa}>$)
(MZM40, MZM42, MZM50, MZM75, MZM84, and MZM85).
The mean $\sigma I_{circa}$ is 0.13 mag.

\begin{figure}
\begin{center}
\resizebox{0.9\hsize}{!}{\includegraphics[angle=0]{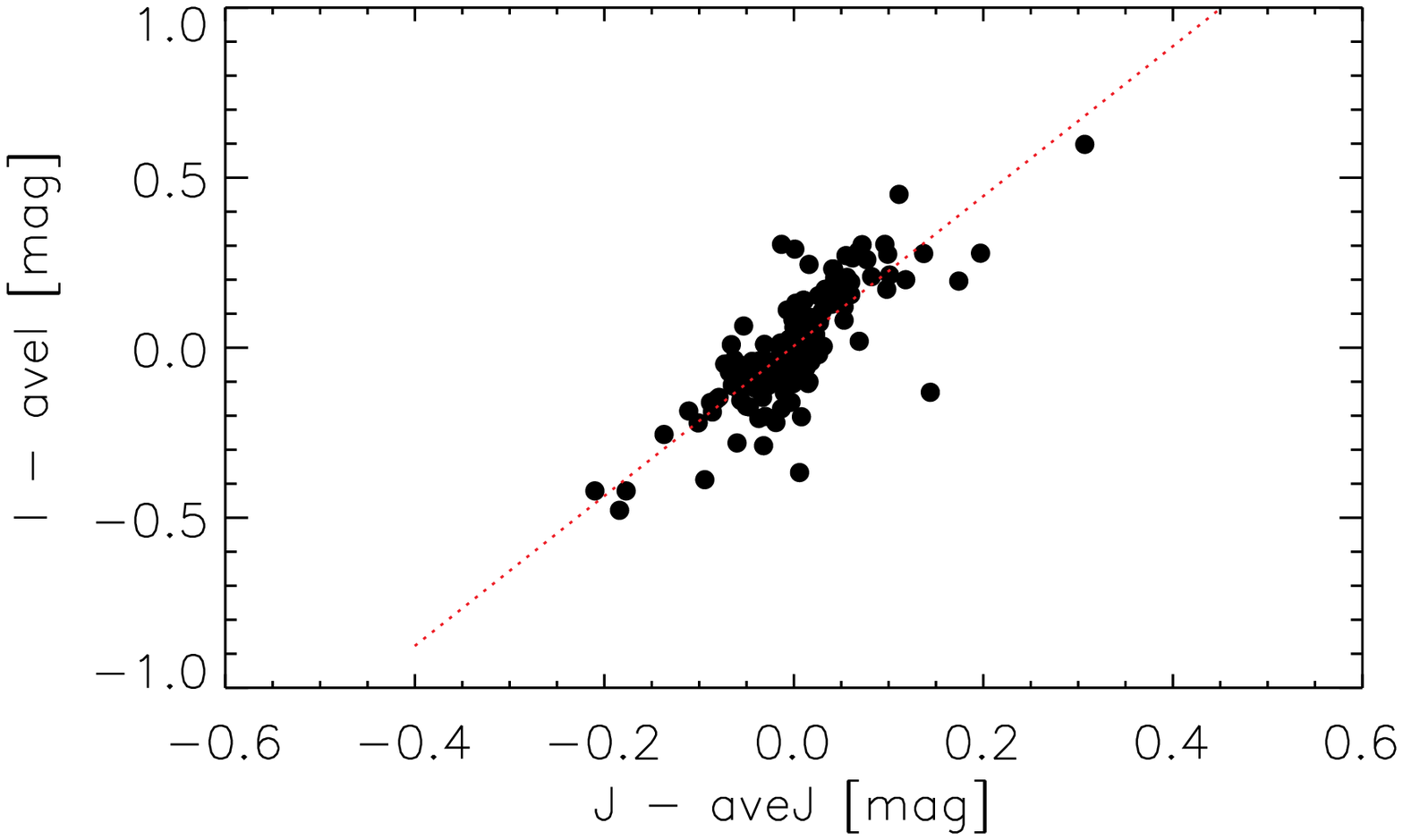}}
\resizebox{0.9\hsize}{!}{\includegraphics[angle=0]{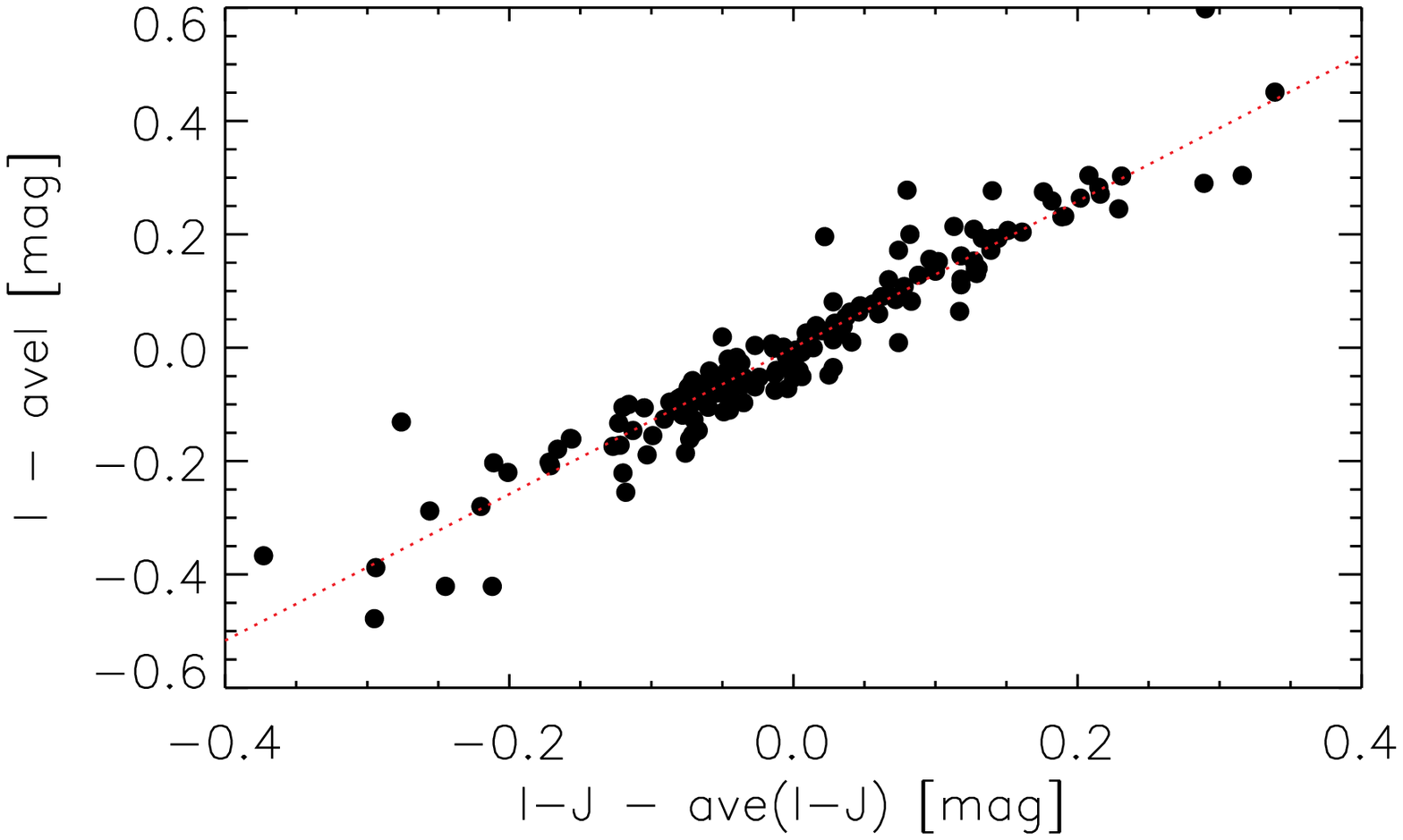}}
\end{center}
\caption{ \label{deltaIdeltaJ} {\it Top panel:} 
Variations in $I$-band vs. those simultaneously measured in $J$-band; 
all  epochs of  targets with detected variability are shown. 
In red a linear fit to the data. 
{\it Bottom panel:} Variations of  $I$ magnitudes vs. those
measured in the $I-J$ colour. In red a linear fit to the data.} 
\end{figure}

Despite the small numbers of  detected variables,
correlated trends appear in  several 
$I$-band  and $J$-band light curves,
as shown in Figs.   \ref{deltaIdeltaJ} and \ref{lighcurve}.
16 targets (out of the 42 detected in both bands) have 
a Pearson correlation coefficient between $I$-band and $J$-band 
measurements above 50\%, and  
the $\sigma_*(I)$ or $\sigma_*(J)$ of their 
$I$-band or $J$-band measurements 
is larger than twice the $\sigma$  of corresponding
field stars with similar magnitudes 
($<$std\_cal(J)$>$ or $<\sigma I_{circa}>$). 
In this latter calculation, 38\% of the sample
shows variations, MZM06, MZM09, MZM10, MZM16, MZM20, MZM22, 
MZM33, MZM34, MZM40, MZM42, MZM50,
MZM59, MZM75, MZM83, MZM84, and MZM85.\\ 
The use of multi-wavelength data improves the detection of 
variables  and the correlated patterns make 
it more solid and reliable.
When using  combined $IJ$-bands,  the  variable detection threshold 
in a single band is lowered 
(from 3.5 $\sigma$ to 2 $\sigma$) to detect more variables.

The Gaia variables listed by Mowlavi et al. (2021) were detected with a detection thresholds of 0.06 mag in $G$ band.
It appears that variables with amplitudes in $G$ band larger than 0.08 mag are all retrieved.
There are eight ANDICAM variables   with 
estimated $G$ variations 
below the 0.06 mag threshold for variability. 
While the Gaia variables retrieved by ANDICAM have average amplitudes in 
$I$-band of 0.27 mag (0.10 mag in $J$-band) and are mostly $> 0.21$ mag,  
variables found by ANDICAM, but not in Mowlavi et al. (2021), 
have average $I$-band amplitudes of 0.16 mag (0.09 in $J$-band) and are mostly $< 0.21$ mag.
We conclude that the detection of variables is complete for $I$-band amplitudes larger than 0.21 mag.

While $G$ band amplitudes are determined for stars with a wide range of $G$ band (from 10 to 20 magnitudes),
the variables reported in AAVSO are brighter than a G magnitude of 14.5.

A number of 24 (out of 42) stars, 
57\% of the sample, have $J$ and $I$ variations 
below the adopted detection threshold (2 $\sigma$). 

The  standard deviations of the targets and calibrators 
are listed in Table \ref{table.mag}.

\begin{figure*}
\begin{center}
\resizebox{0.9\hsize}{!}{\includegraphics[angle=0]{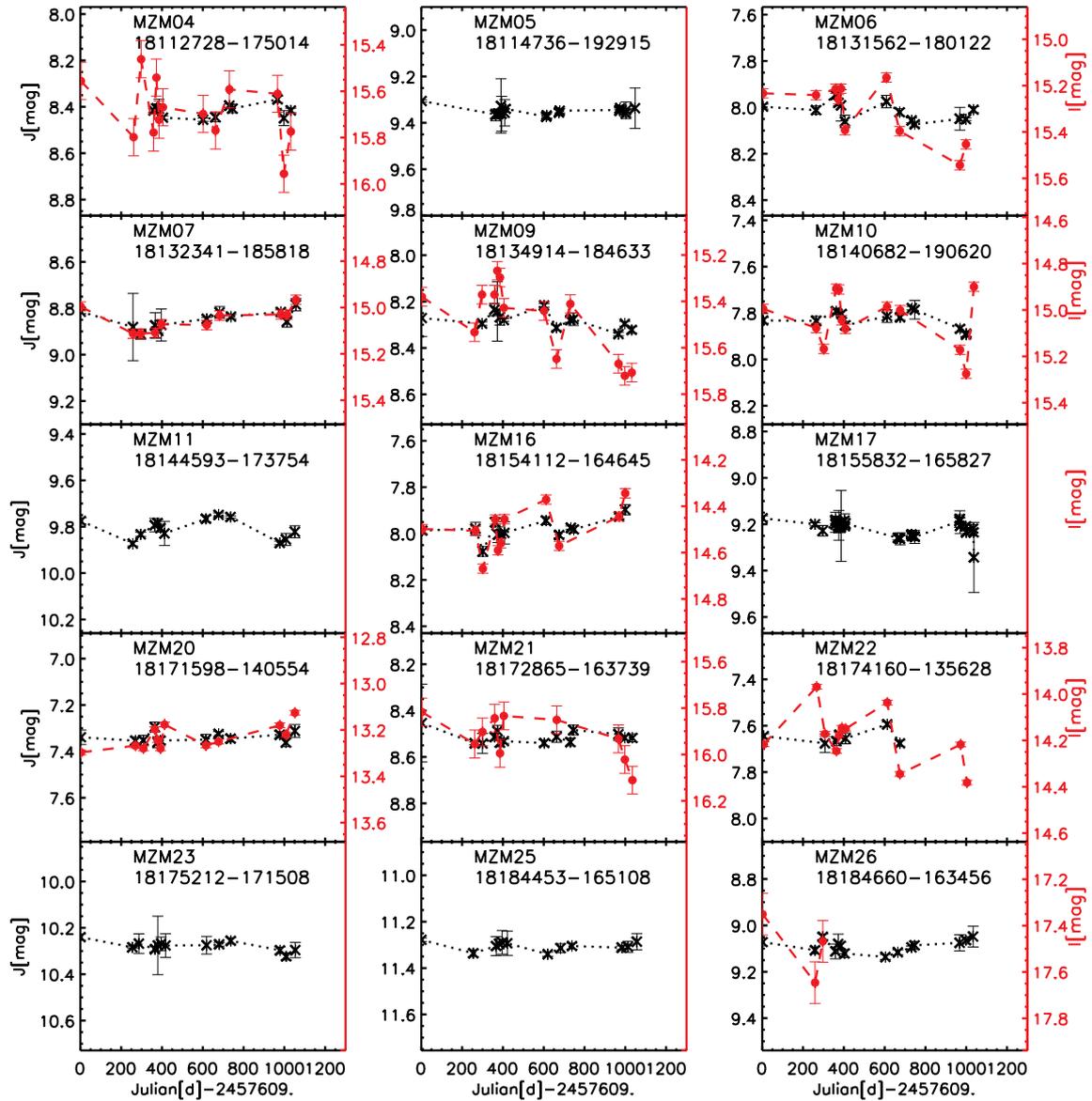}}
\end{center}
\caption{\label{lighcurve} 
In each panel,  on the left y-axis, the $J$-band magnitudes 
vs. Julian days of one observed star are plotted with black X.
Dotted-black lines connect the X points.
The $I$-band magnitudes are over-plotted 
with red-filled circles and their values annotated on the right y-axis. 
Long-dashed red lines connect the circles. 
Two horizontal dotted-dashed lines are drawn at  $\pm0.15$ mag distance
from the average magnitudes of the observed star.
} 
\end{figure*}

\addtocounter{figure}{-1}
\begin{figure*}
\begin{center}
\resizebox{0.9\hsize}{!}{\includegraphics[angle=0]{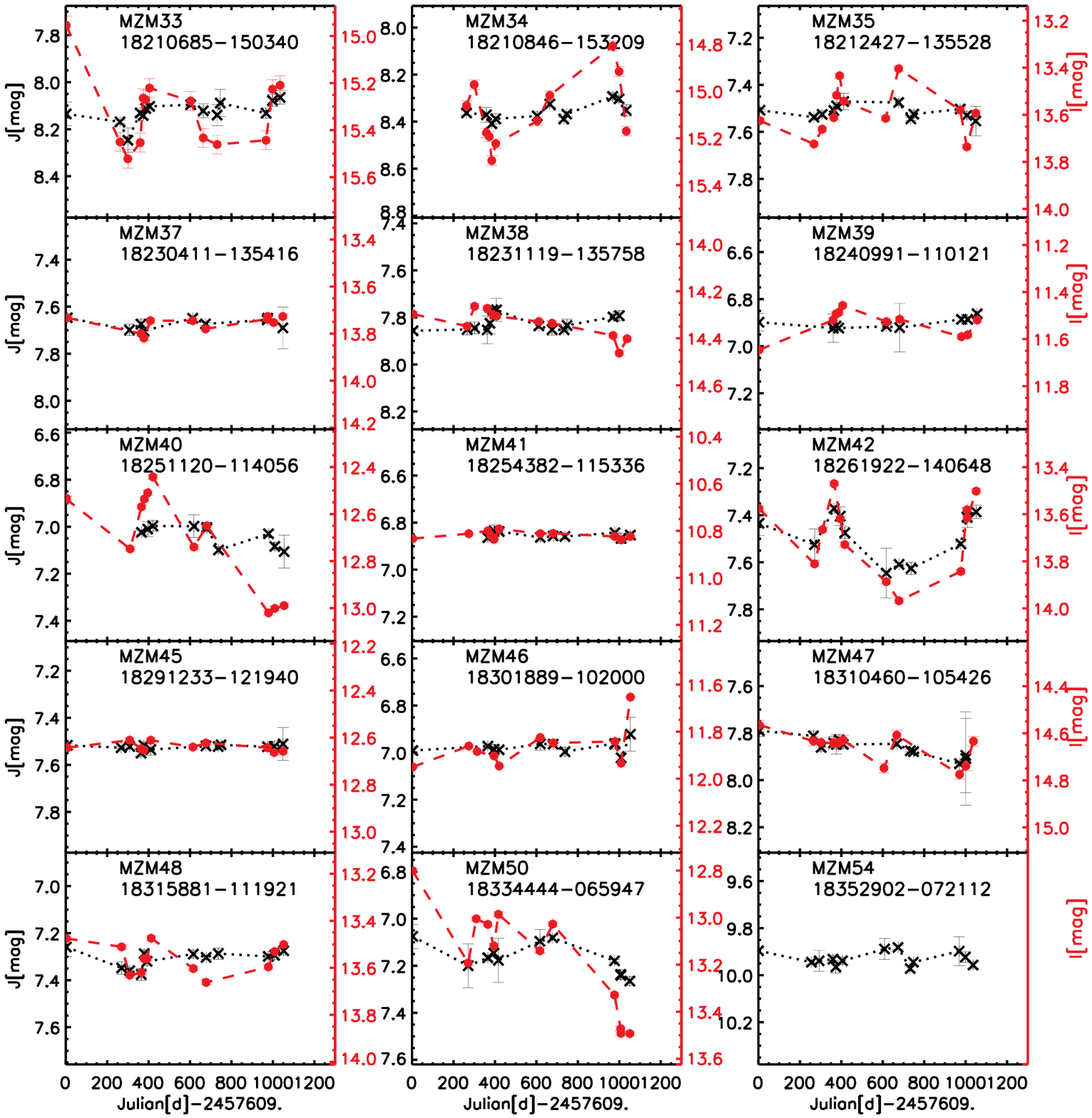}}
\end{center}
\caption{Continuation of Fig.\ \ref{lighcurve}. } 
\end{figure*}

\addtocounter{figure}{-1}
\begin{figure*}
\begin{center}
\resizebox{0.9\hsize}{!}{\includegraphics[angle=0]{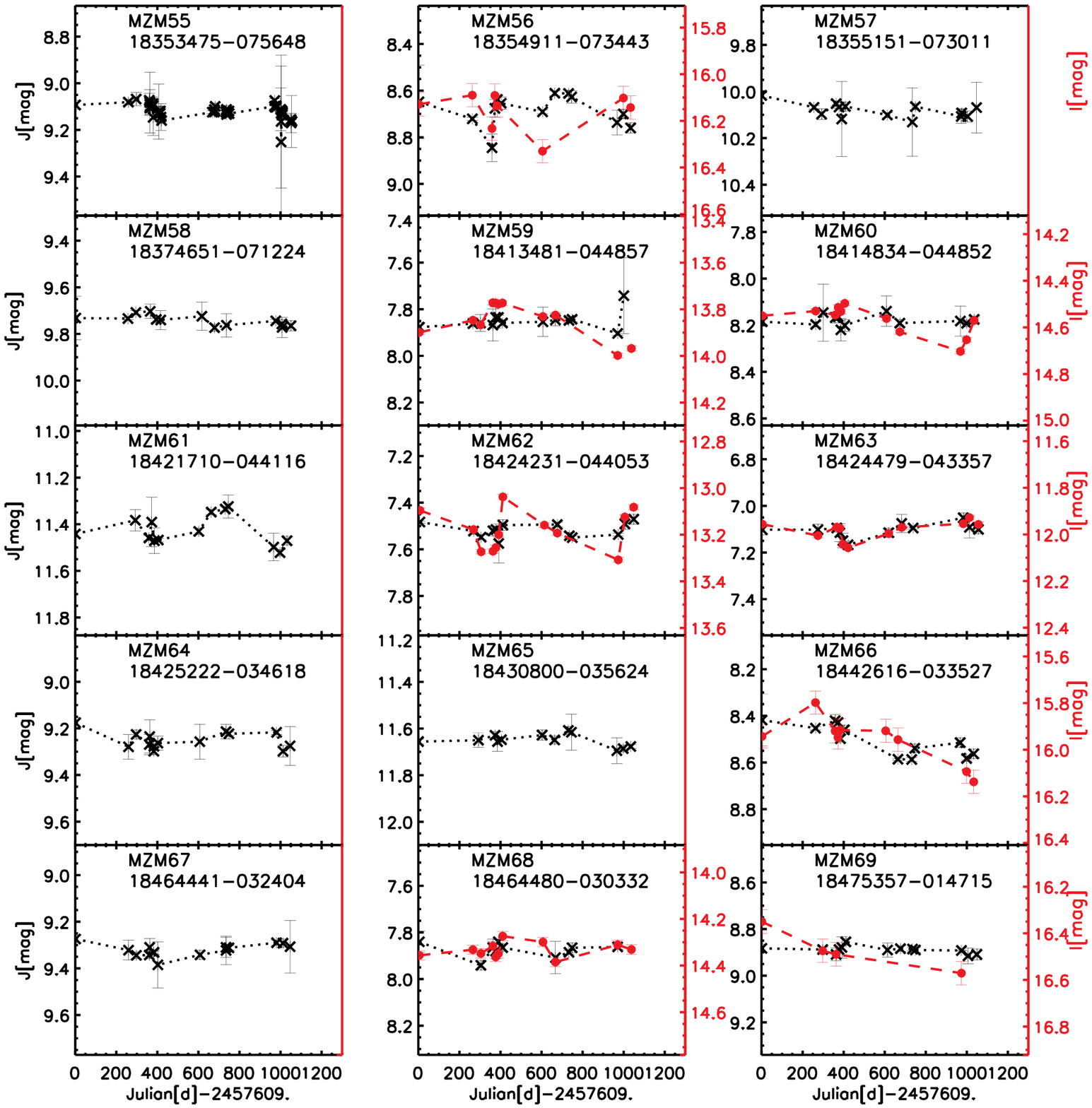}}
\end{center}
\caption{Continuation of Fig.\ \ref{lighcurve}. } 
\end{figure*}

\addtocounter{figure}{-1}
\begin{figure*}
\begin{center}
\resizebox{0.9\hsize}{!}{\includegraphics[angle=0]{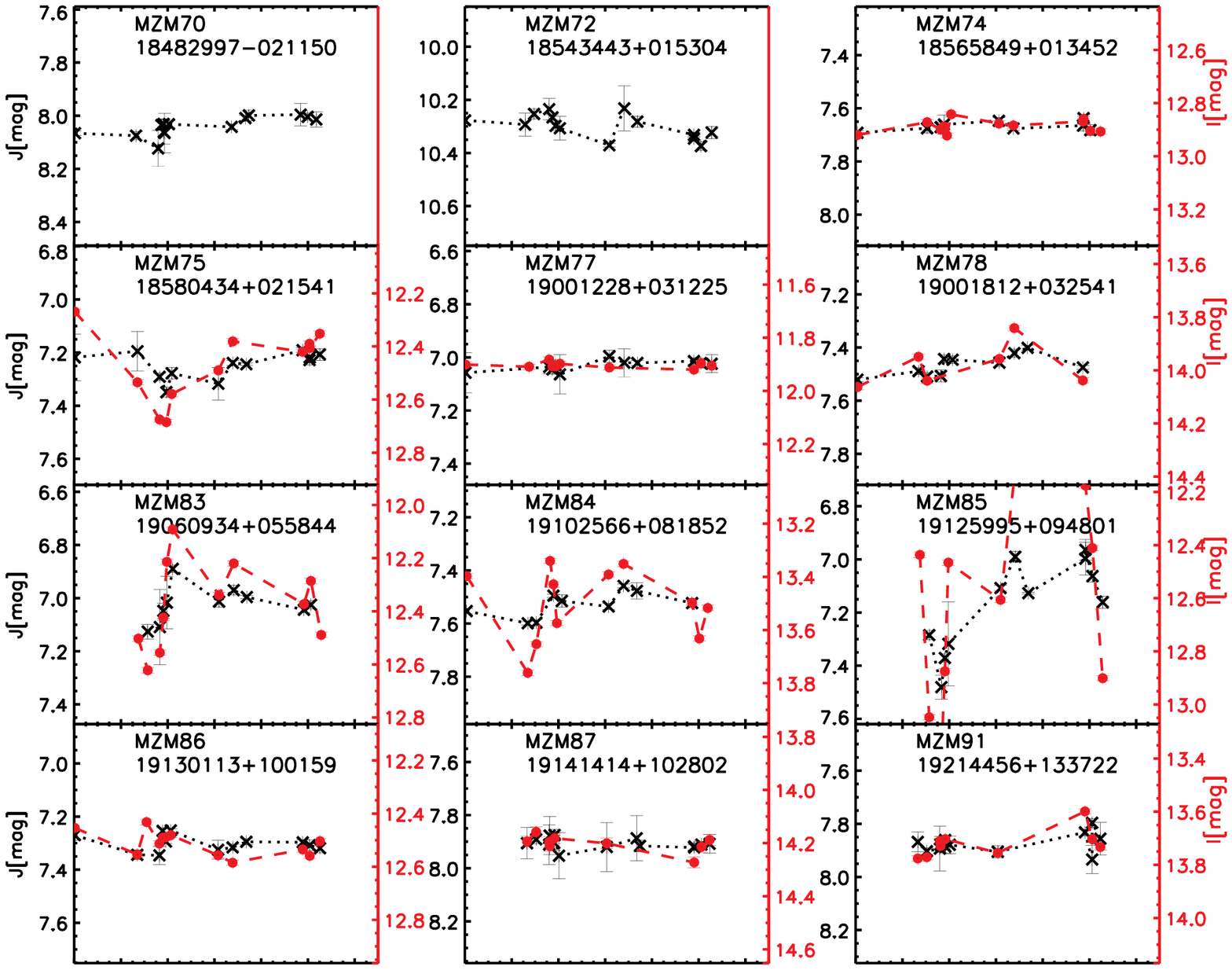}}
\end{center}
\caption{Continuation of Fig.\ \ref{lighcurve}. } 
\end{figure*}

\subsection{Peridiograms}
Periods are obtained as the highest power 
in the Lomb-Scargle periodogram, 
which is an elaborated version of the classical
peridiogram for unevenly sampled data \citep{scargle82,horne86}.
The scargle.pro routine in the NASA IDL Astronomy User’s 
Library was used \citep{landsman93}. 
The analysis of the $I$-band Lomb-Scargle peridiogram 
yields periods for 8 stars (out of the  16  variables in Sect. \ref{significant}),
which are listed in Table \ref{mag-compare-target}.
Only periods with power levels corresponding 
to false alarm probabilities (the probability 
for the periods to be false)  below 50\% 
were considered.   
The power of MZM83 corresponds to a false alarm probability (FAP) of 
43.6\%, while the probability remains below 25\% for the other 7 stars.
Their periods  range from 167.6-433.3 days.

A higher FAP brings a better census of variables; however, 
it may bring false cases. A FAP value of 10\% is commonly used 
in time-series photometry, as mentioned in the work of 
\citet{vanderplas18}. FAP values of 50\% are also found in literature, 
for example in the spectroscopic 
time-series analysis of  \citet{cincunegui07}.
Besides the Scargle method, a fitting of the light curve  
was performed and  the sinusoidal curves are evident.

The 8 periodograms and phased light curves are
shown in Appendix.
In conclusion, 38\% of the 42 targets with ANDICAM $IJ$ detections
are found to be variable, and 19\% show periodicity.

\subsection{Amplitudes of the variations} 
16 late-type stars (out of 42 detected in both $IJ$ bands) 
appear to have  significant light variations when compared
with surrounding field stars, as described 
in Sect. \ref{significant}. 
The measured variations range from 0.038-1.076 mag
in $I$-band, and from 0.031-0.517 mag in $J$-band,
and are listed in Table \ref{table.mag}.  
As  mentioned in Sect. \ref{significant},
the variations in $I$ and $J$ bands appear correlated.
For  pairs of simultaneously 
taken $I$ and $J$ magnitudes of the 16 variables,  
the $I$ variations, $\Delta I$, are plotted  against 
the $J$  variations, $\Delta J$, 
in Fig.\ \ref{deltaIdeltaJ}.
The  $\Delta I$ values are $2.2 \pm 0.1$ times larger than 
the $\Delta J$ values, and $1.292\pm0.004$ larger than those
in the $I-J$ colors, $\Delta (I-J)$.

For the eight stars found to be periodic variables, 
the least-squares fitting of data   by   sinusoidal curves of the type 
${\rm I(t)} = \frac{\rm \Delta I}{2} sin[2 \pi (\frac{\rm t}{\rm Per} + \omega _{\rm o} )] + {\rm <I>}$
 was performed. For each curve, three parameters were estimated;  
$\frac{\Delta I}{2}$ is the semi-amplitude of the pulsation
(i.e., the difference in absolute value 
between the mean value and the minimum or maximum
deviation),
$\omega _o$ is the value at zero phase (at the maximum), 
and $<I>$ is the mean
magnitude of the pulsator. 
The $\frac{\Delta I}{2}$ values are listed in 
Table \ref{mag-compare-target},
and shown in Appendix.

\section{Previously known variables}

For nine targets, amplitudes are reported in 
the AAVSO International Variable Star database (VSX) 
\citep{AAVSO}, four of which were not detected as significantly
varying in this work. They range from 0.10-0.64 mag.\\
In the work of  \citet{mowlavi21},  
there are estimates of  G-band 
amplitudes for  15 of the ANDICAM 
targets (10 of which were classified as a variable in this work);
they are based on the G-band photometric errors 
and range from 0.06-0.39 mag. These stars
are not listed in the GAIA DR2 table of   
long-period variables (LPV). 
This brings to 59\% the fraction of variable targets and
known amplitudes.

For seven targets, periods from 109-457 d are listed  
in the International Variable Star Index VSX 
\citep{watson06}, four of which are not detected as 
significantly varying in this work.  
MZM83 has a period of 148.5 d in the AAVSO catalog,
in the ANDICAM data a low power peak appears  at 279 d with a
high probability of 43\% to be false. 
The other six AAVSO stars have no periods detected in ANDICAM.

This increases from 19\% to 33\% the number of targets with 
known periods.

\section{\Mk\ versus periods of known Galactic RSGs}

\begin{figure*}
\begin{center}
\resizebox{0.8\hsize}{!}{\includegraphics[angle=0]{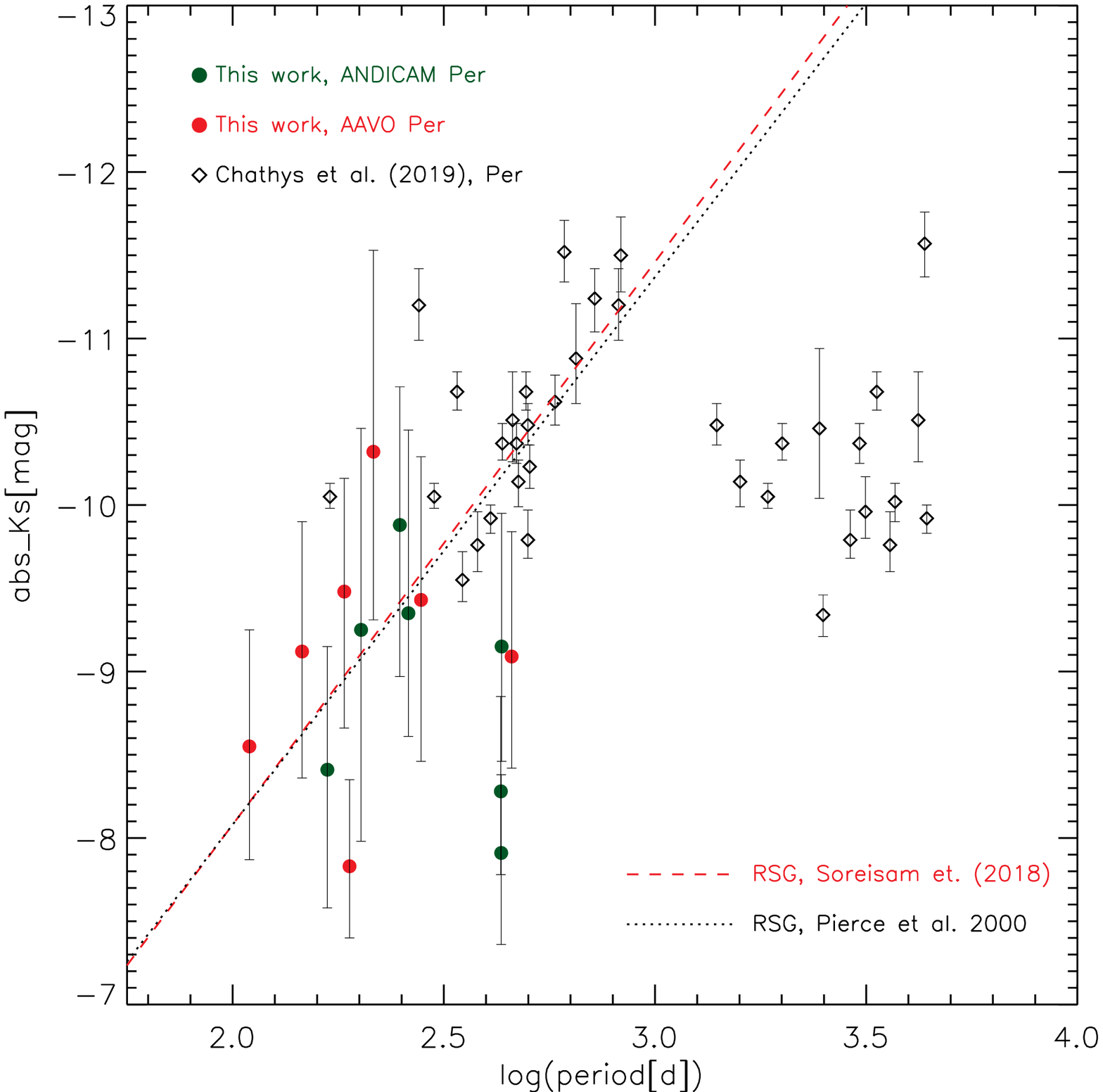}}
\end{center}
\caption{ \label{chathys.eps} 
 \Mk\ values vs. periods of Galactic RSGs (black diamonds). 
 RSG periods are from \citet{chatys19} and 
\Mk\ magnitudes based on Gaia EDR3 distance are from 
\citet{messineo19} and \citet{messineo21z}. 
Late-type from this work  are overplotted 
marked with filled circles in red when the periods 
are taken from the AAVSO catalog and in dark-green when 
the periods  are determined with ANDICAM data. The distances 
are from Gaia EDR3 \citet{messineo21z}.
}
\end{figure*}

To verify the newly obtained periods, the \Mk\ versus period diagram
of the targets is compared with that of well-studied variable RSGs
in Fig.\ \ref{chathys.eps}.
For known variables, the periods are taken from 
\citet{chatys19} and the \Mk\ values 
are those calculated by \cite{messineo21z} with EDR3 
Gaia distances and are  obtained as 
described in \citet{messineo19}.
The distances adopted are based on EDR3 Gaia parallaxes and 
are the  geometric distances by \citet{bailer21}.
The short periods ($< 2,000 $ d) appear to describe a
clear sequence in this plane. The sequence appears much improved
with the use of  EDR3 Gaia distances, and has  a   $\sigma=0.32$ mag.
The sequence is consistent with the fit made for RSGs in  
Perseus OB1 by \citet{pierce00}, as well as with the fit
obtained for RSGs in M31 by \citet{soraisam18}.

At this stage, in the \Mk-Period plane, 
the  distribution of the 14 ANDICAM targets with 
determined periods appears
consistent with that of known RSGs, within errors.

\section{Summary  and remarks}
ANDICAM observations of a sample of late-type stars 
were obtained  over  a 
 1054 day time period (2.9 years).
57 bright late-type targets from the sample of \citet{messineo17} 
were observed in  $J$ bands and 42 were 
detected in the simultaneously taken $I$ band images.
It appears that at least 38\% 
of the ANDICAM sample is made of variable stars,
47\% when including additional information in AAVSO,  
or 59\% when considering
also the Gaia  amplitude estimates reported by \citet{mowlavi21}. 
Furthermore, 19\% have detected periodic behaviours in the ANDICAM data 
and  33\% when including the AAVSO periods. 

The targeted late-type stars have average ANDICAM $<J>$ 
from 6.85-11.65 mag and ANDICAM $<I>$ from 10.82-17.49 mag. 
However, despite their faintness
in $I$-band, $I$-band is more suitable than $J$-band
for detecting variables.
Indeed, the magnitude variations measured in  $I$-band
are correlated with those seen in $J$-band and
are  a factor 2.2 larger. 
In $I$-band, the differences between the minimum and 
the maximum magnitudes of each star range from 0.04-1.08 mag, 
while in $J$-band from 0.03-0.52 mag.\\
The ANDICAM data here presented indicates that
the $I$ magnitudes and $I-J$ colors of the targets
are varying in a correlated manner and that
$\Delta I \propto  1.29 \Delta (I-J)$.
In pulsating large-amplitude stars, the  amplitudes 
are known to decrease with increasing wavelength, 
as the envelope expands and cools down every radial 
pulse \citep{reid02}. For example, in Mira stars 
the $V$-band amplitudes can be 
even 8 mag \citep{reid02},
and the $J$-band amplitudes  are about  1.6 times 
larger than those measured in \Ks\ band \citep{messineo04}. 
This effect is caused by   a changing opacity in the stellar 
atmosphere \citep{reid02}, due to   molecular bands.
When a late-type star pulses, it expands and cools down,
and it  reaches its minimum light when the atmospheric opacity 
is at the maximum value \citep{reid02}.
When a long-period variable pulses, it cyclically changes its spectral type. 
As  mentioned by \citet{pierce00}, for RSGs the measured light variations  are found to be 
a function of the used filter. The  $I$-band spectrum is dominated by TiO molecular bands which are 
extremely sensitive to  temperature variations, while the $J$-band spectrum is not affected by TiO bands.
The observed correlation between the ANDICAM color variations, $\Delta (I-J)$, and the magnitude variations 
are, therefore, consistent with the expected behaviour for radial pulsation. 
For normal (static) giants and RSGs,  
a  change of spectral type from M2 to M5 (M3.5 $\pm 1.5$)  corresponds to  a  color change 
$\Delta (I-J) \approx 0.26$ mag \citep{johnson66}. Variable RSGs could have larger color changes,
due to their larger radii and the large convective cells present in their turbulent atmospheres.

As shown by \citet{kiss06} and \citet{chatys19}, variable RSGs are of late-type (M-type).
A broad correlation is found between the stellar luminosity and the $R$-band amplitudes of 120 RSGs
\citep{soraisam18}. In turn, as  average stellar temperatures decrease with increasing luminosity, 
a positive correlation also exists between the stellar temperatures and amplitudes 
\citep{messineo19,messineo20var}. 
The sample here analyzed is small and no clear trend is observed between the luminosity 
and  amplitudes. However, the small amplitudes are in agreement with supergiant classification.
Indeed, the most luminous AGB stars (e.g., super-AGBs) 
are expected to be large-amplitude pulsators
\citep{oGrady21}. 

New periods are determined for eight late-type stars and range
from 167-433 d, and seven periods are available from AAVSO.
The time baseline of the ANDICAM data does not allow  us to check for
long secondary periods (LSP), which are typically 
longer than $2,000$ d. In the Galaxy, LSP periods are seen in 50\% of 
the RSGs  \citep{kiss06,chatys19}.

The sample does not contain large-amplitude variables, which are
bright luminous AGBs.
For the targets found to be periodic variables, their distribution in the 
period-luminosity diagram suggests that they are RSGs, consistently
with the work of \citet{messineo17}.
However, distance errors are still large and
it is better to re-check with the final Gaia parallaxes. 
Gaia will also release  spectra and G-band light curves.

Due to the significant  uncertainties in their distances,
time-series measurements  appear to offer a promising 
means of assessing the stellar luminosity class of 
obscured inner Galactic objects.

\begin{appendix}
\section{Phased light curves}
In Fig.\ \ref{fig.peridiograms}, the phased light curves are shown.
Periods are estimated for eight variables.

\begin{figure*}
\begin{center}
\resizebox{0.45\hsize}{!}{\includegraphics[angle=0]{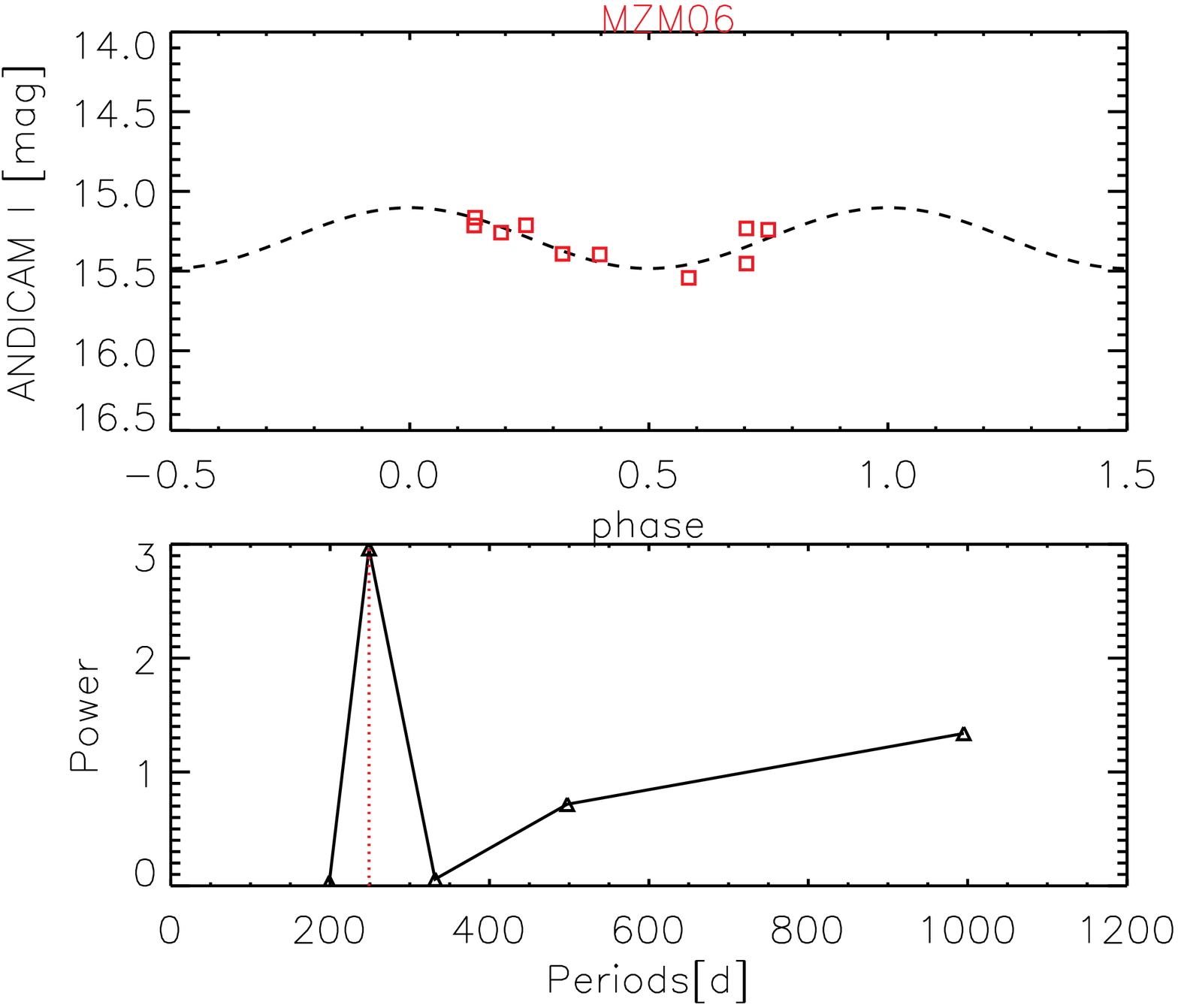}}
\resizebox{0.45\hsize}{!}{\includegraphics[angle=0]{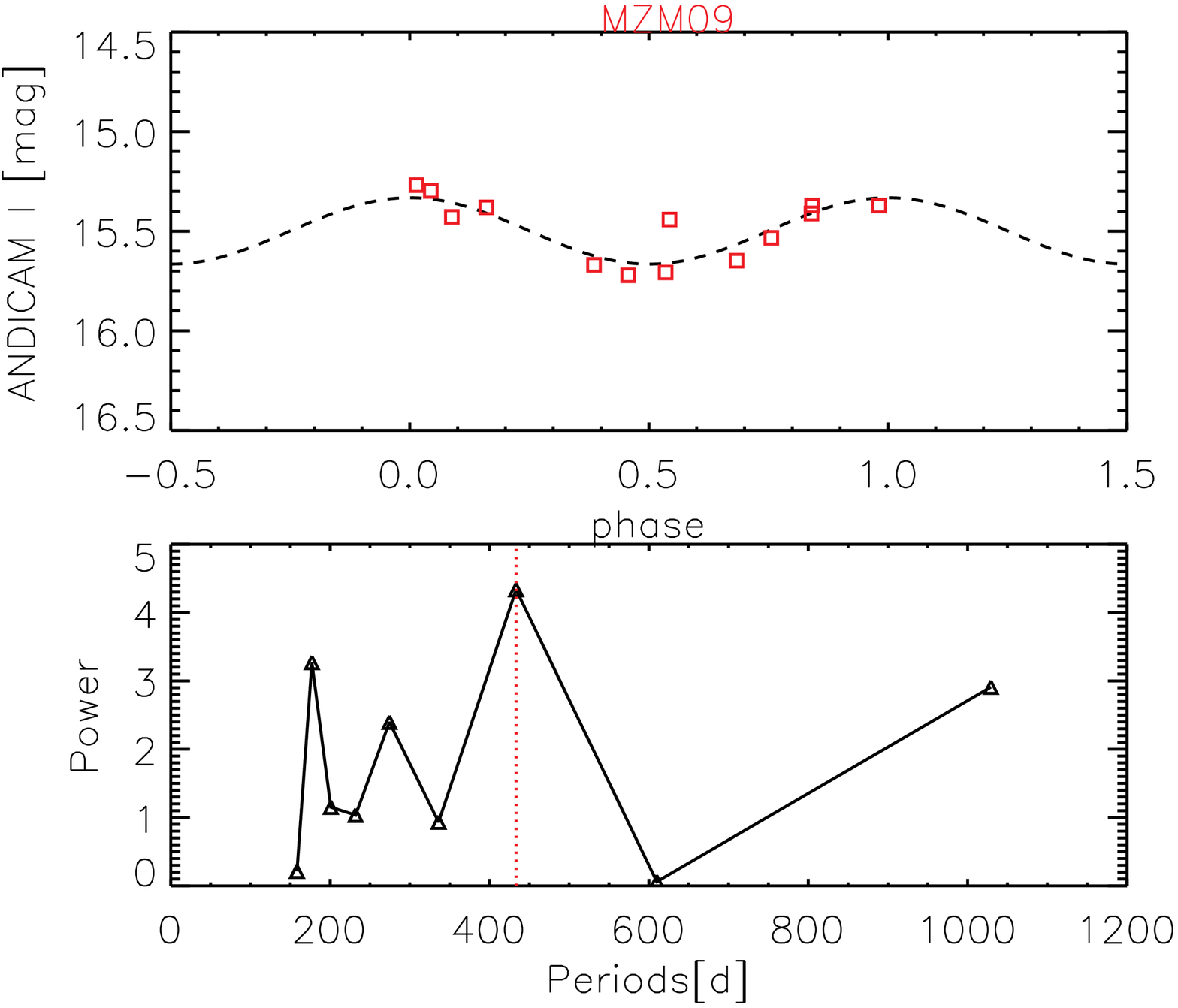}}
\resizebox{0.45\hsize}{!}{\includegraphics[angle=0]{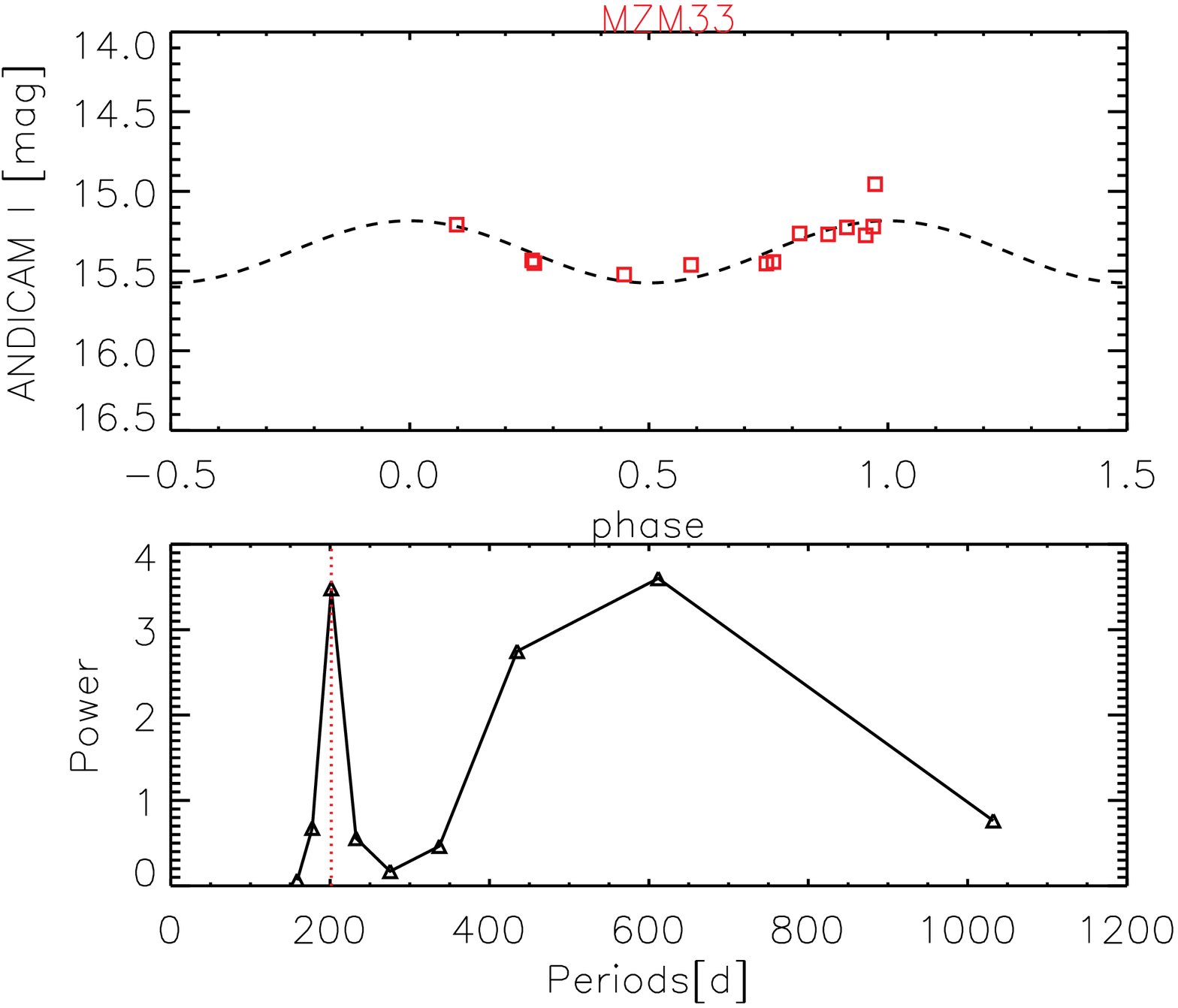}}
\resizebox{0.45\hsize}{!}{\includegraphics[angle=0]{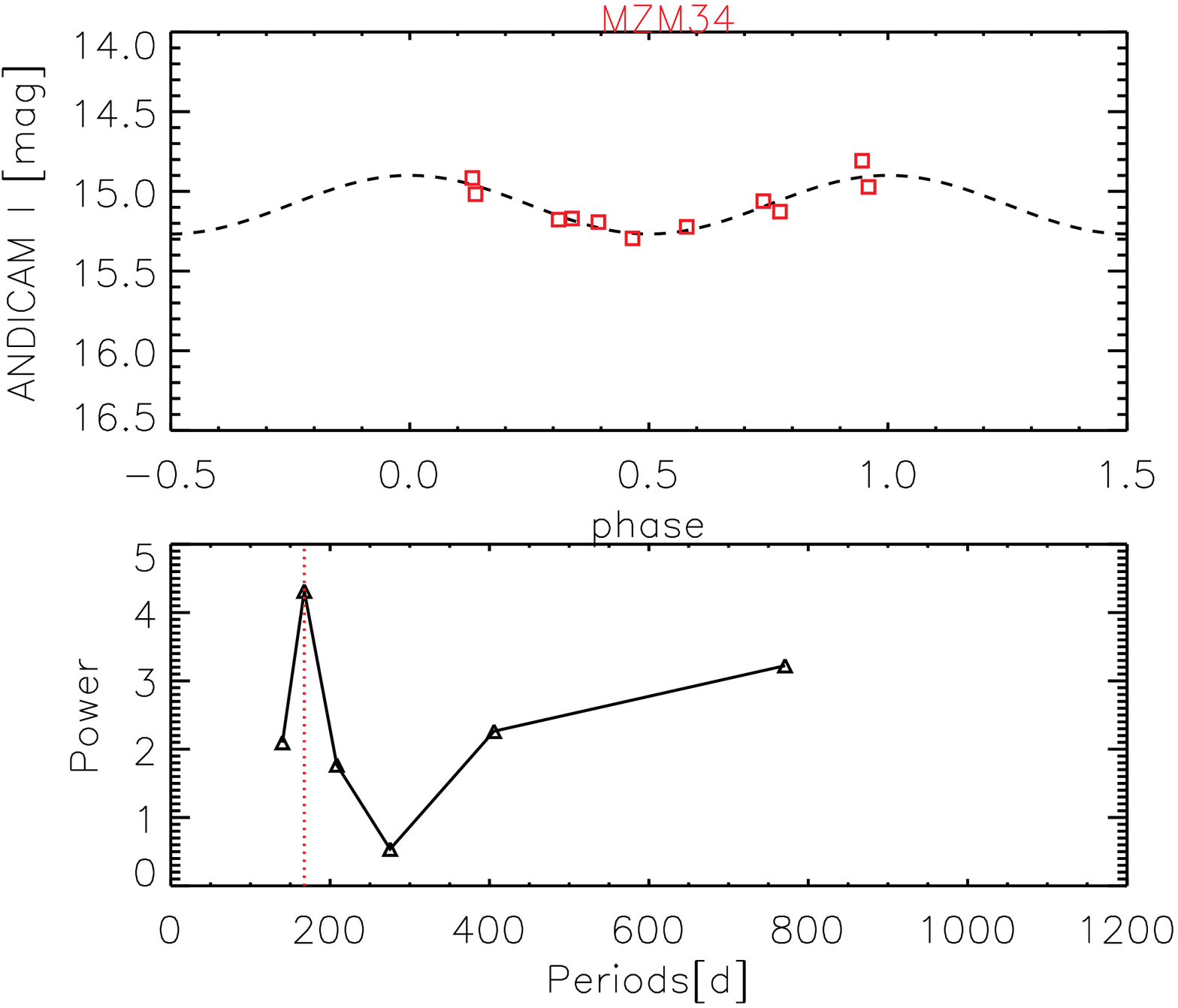}}
\end{center}
\caption{\label{fig.peridiograms} 
For  variables with an identified periodicity, {\it in the top panel}
the sinusoidal curve vs. the phase is shown, and {\it in the bottom panel}
the Lomb-Scargle periodogram. The red vertical dotted line marks the
position of the adopted period. }  
\end{figure*}

\addtocounter{figure}{-1}
\begin{figure*}
\begin{center}
\resizebox{0.45\hsize}{!}{\includegraphics[angle=0]{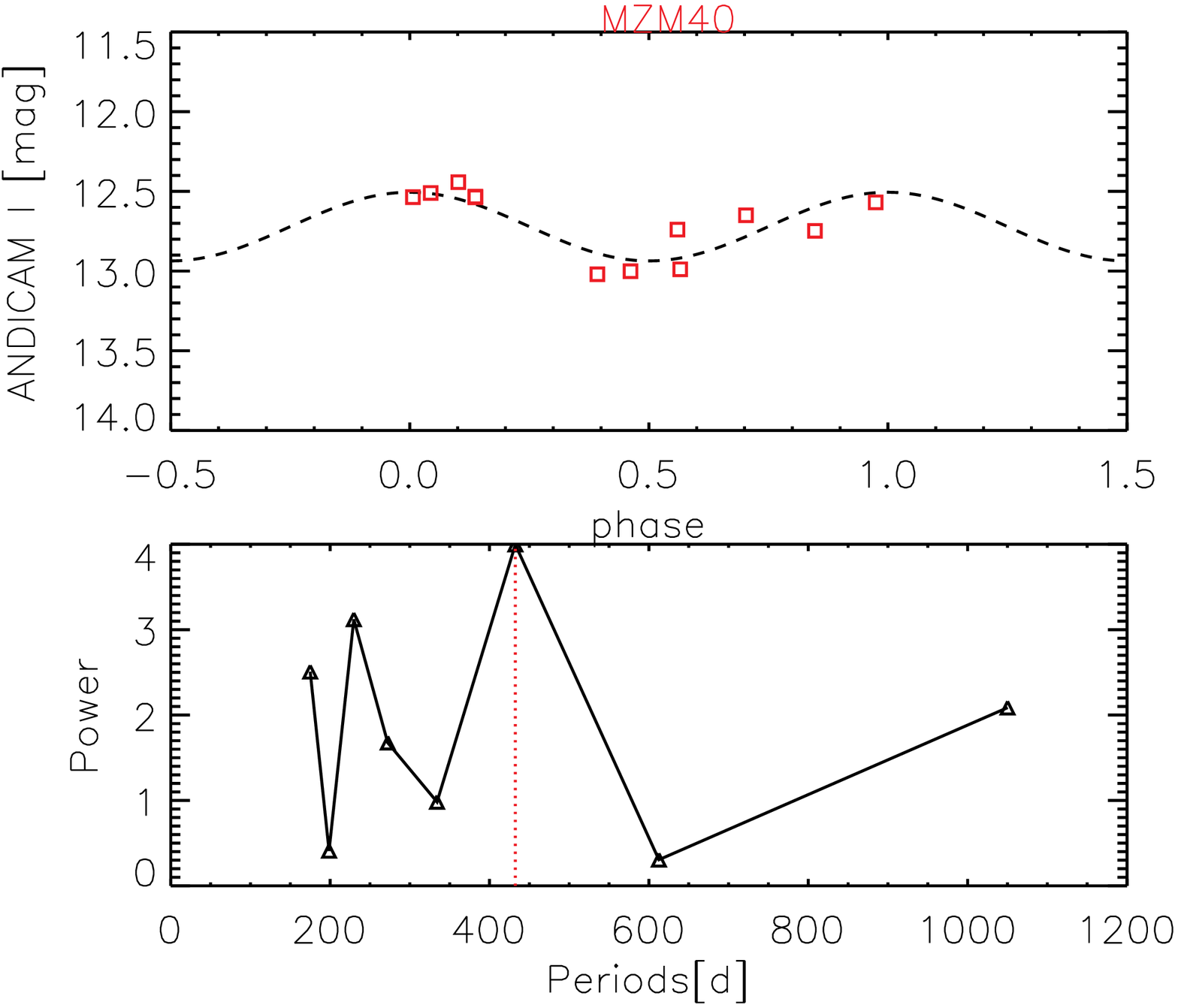}}
\resizebox{0.45\hsize}{!}{\includegraphics[angle=0]{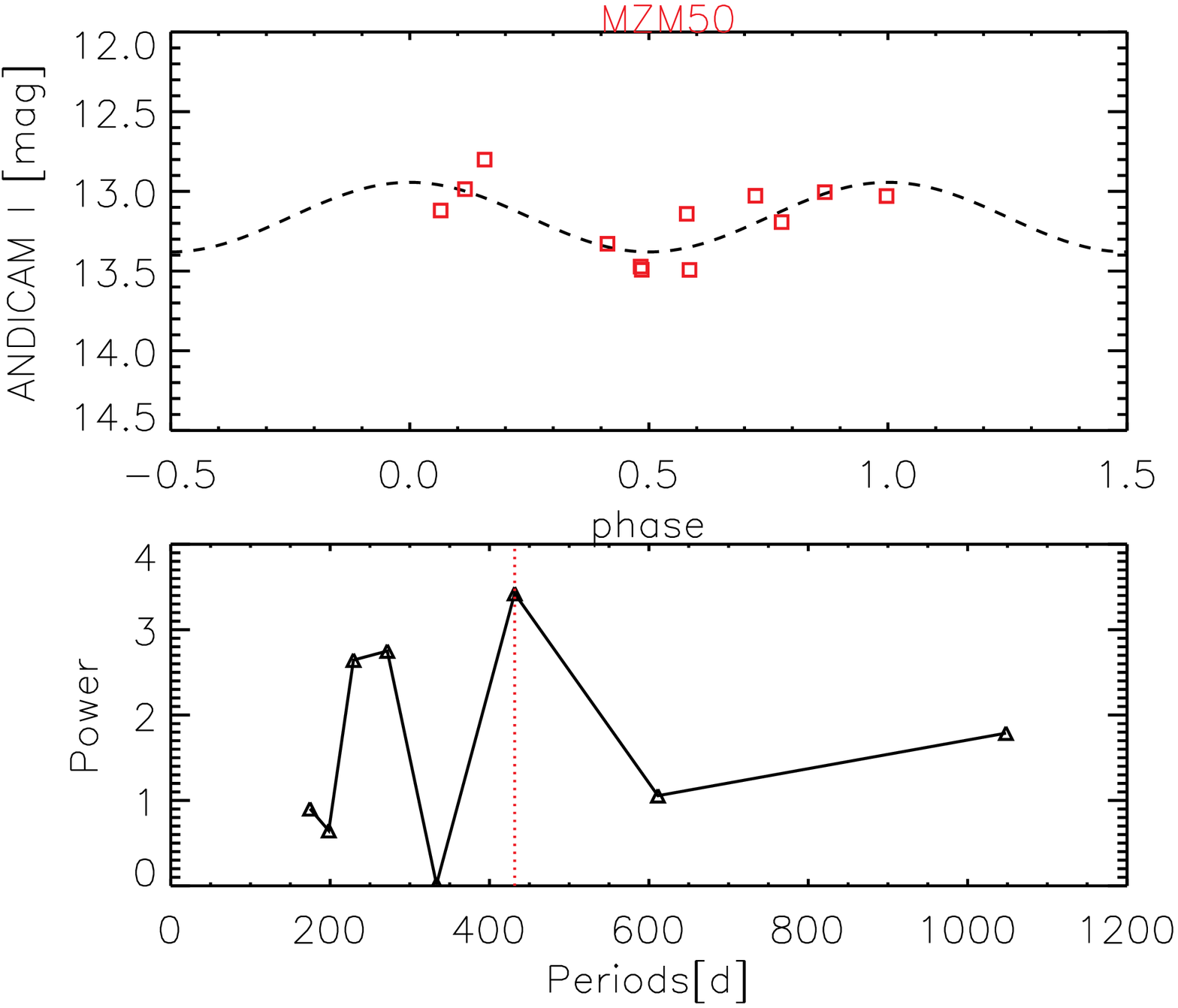}}
\resizebox{0.45\hsize}{!}{\includegraphics[angle=0]{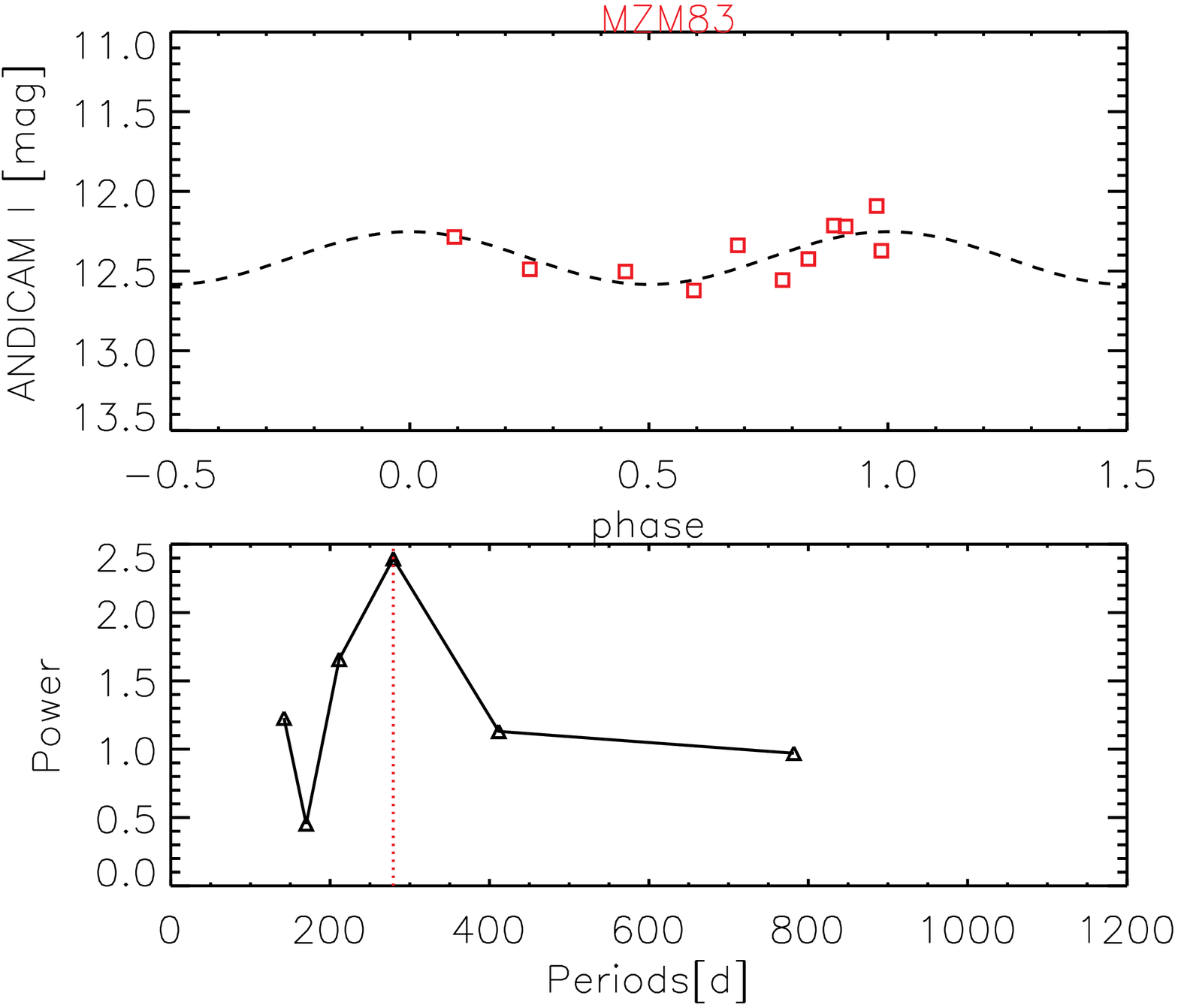}}
\resizebox{0.45\hsize}{!}{\includegraphics[angle=0]{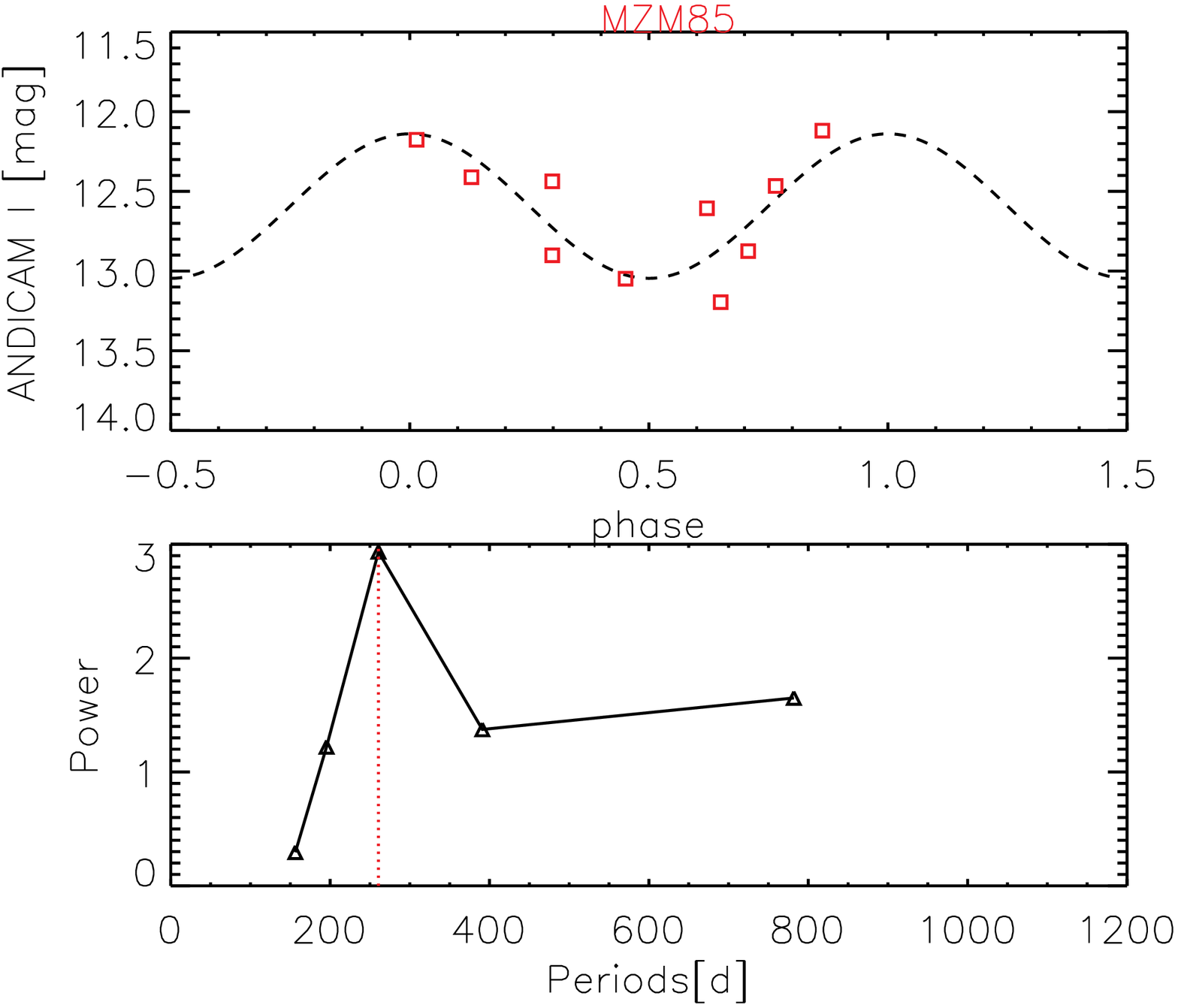}}
\end{center}
\caption{Continuation of Fig.\ \ref{fig.peridiograms}.}
\end{figure*}

\end{appendix}

%\bibliographystyle{aa}
%\bibliography{biblio}

\begin{thebibliography}{44}
\expandafter\ifx\csname natexlab\endcsname\relax\def\natexlab#1{#1}\fi

\bibitem[{{Alam} {et~al.}(2015){Alam}, {Albareti}, {Allende Prieto}, {Anders},
  {Anderson}, {Anderton}, {Andrews}, {Armengaud}, {Aubourg}, {Bailey}, {Basu},
  {Bautista}, {Beaton}, {Beers}, {Bender}, {Berlind}, {Beutler}, {Bhardwaj},
  {Bird}, {Bizyaev}, {Blake}, {Blanton}, {Blomqvist}, {Bochanski}, {Bolton},
  {Bovy}, {Shelden Bradley}, {Brandt}, {Brauer}, {Brinkmann}, {Brown},
  {Brownstein}, {Burden}, {Burtin}, {Busca}, {Cai}, {Capozzi}, {Carnero
  Rosell}, {Carr}, {Carrera}, {Chambers}, {Chaplin}, {Chen}, {Chiappini},
  {Chojnowski}, {Chuang}, {Clerc}, {Comparat}, {Covey}, {Croft}, {Cuesta},
  {Cunha}, {da Costa}, {Da Rio}, {Davenport}, {Dawson}, {De Lee}, {Delubac},
  {Deshpande}, {Dhital}, {Dutra-Ferreira}, {Dwelly}, {Ealet}, {Ebelke},
  {Edmondson}, {Eisenstein}, {Ellsworth}, {Elsworth}, {Epstein}, {Eracleous},
  {Escoffier}, {Esposito}, {Evans}, {Fan}, {Fern{\'a}ndez-Alvar}, {Feuillet},
  {Filiz Ak}, {Finley}, {Finoguenov}, {Flaherty}, {Fleming}, {Font-Ribera},
  {Foster}, {Frinchaboy}, {Galbraith-Frew}, {Garc{\'\i}a},
  {Garc{\'\i}a-Hern{\'a}ndez}, {Garc{\'\i}a P{\'e}rez}, {Gaulme}, {Ge},
  {G{\'e}nova-Santos}, {Georgakakis}, {Ghezzi}, {Gillespie}, {Girardi},
  {Goddard}, {Gontcho}, {Gonz{\'a}lez Hern{\'a}ndez}, {Grebel}, {Green},
  {Grieb}, {Grieves}, {Gunn}, {Guo}, {Harding}, {Hasselquist}, {Hawley},
  {Hayden}, {Hearty}, {Hekker}, {Ho}, {Hogg}, {Holley-Bockelmann}, {Holtzman},
  {Honscheid}, {Huber}, {Huehnerhoff}, {Ivans}, {Jiang}, {Johnson},
  {Kinemuchi}, {Kirkby}, {Kitaura}, {Klaene}, {Knapp}, {Kneib}, {Koenig},
  {Lam}, {Lan}, {Lang}, {Laurent}, {Le Goff}, {Leauthaud}, {Lee}, {Lee},
  {Licquia}, {Liu}, {Long}, {L{\'o}pez-Corredoira}, {Lorenzo-Oliveira},
  {Lucatello}, {Lundgren}, {Lupton}, {Mack}, {Mahadevan}, {Maia}, {Majewski},
  {Malanushenko}, {Malanushenko}, {Manchado}, {Manera}, {Mao}, {Maraston},
  {Marchwinski}, {Margala}, {Martell}, {Martig}, {Masters}, {Mathur},
  {McBride}, {McGehee}, {McGreer}, {McMahon}, {M{\'e}nard}, {Menzel},
  {Merloni}, {M{\'e}sz{\'a}ros}, {Miller}, {Miralda-Escud{\'e}}, {Miyatake},
  {Montero-Dorta}, {More}, {Morganson}, {Morice-Atkinson}, {Morrison},
  {Mosser}, {Muna}, {Myers}, {Nandra}, {Newman}, {Neyrinck}, {Nguyen},
  {Nichol}, {Nidever}, {Noterdaeme}, {Nuza}, {O'Connell}, {O'Connell},
  {O'Connell}, {Ogando}, {Olmstead}, {Oravetz}, {Oravetz}, {Osumi}, {Owen},
  {Padgett}, {Padmanabhan}, {Paegert}, {Palanque-Delabrouille}, {Pan},
  {Parejko}, {P{\^a}ris}, {Park}, {Pattarakijwanich}, {Pellejero-Ibanez},
  {Pepper}, {Percival}, {P{\'e}rez-Fournon}, {P{\'e}rez-R{\`a}fols},
  {Petitjean}, {Pieri}, {Pinsonneault}, {Porto de Mello}, {Prada}, {Prakash},
  {Price-Whelan}, {Protopapas}, {Raddick}, {Rahman}, {Reid}, {Rich}, {Rix},
  {Robin}, {Rockosi}, {Rodrigues}, {Rodr{\'\i}guez-Torres}, {Roe}, {Ross},
  {Ross}, {Rossi}, {Ruan}, {Rubi{\~n}o-Mart{\'\i}n}, {Rykoff},
  {Salazar-Albornoz}, {Salvato}, {Samushia}, {S{\'a}nchez}, {Santiago},
  {Sayres}, {Schiavon}, {Schlegel}, {Schmidt}, {Schneider}, {Schultheis},
  {Schwope}, {Sc{\'o}ccola}, {Scott}, {Sellgren}, {Seo}, {Serenelli}, {Shane},
  {Shen}, {Shetrone}, {Shu}, {Silva Aguirre}, {Sivarani}, {Skrutskie},
  {Slosar}, {Smith}, {Sobreira}, {Souto}, {Stassun}, {Steinmetz}, {Stello},
  {Strauss}, {Streblyanska}, {Suzuki}, {Swanson}, {Tan}, {Tayar}, {Terrien},
  {Thakar}, {Thomas}, {Thomas}, {Thompson}, {Tinker}, {Tojeiro}, {Troup},
  {Vargas-Maga{\~n}a}, {Vazquez}, {Verde}, {Viel}, {Vogt}, {Wake}, {Wang},
  {Weaver}, {Weinberg}, {Weiner}, {White}, {Wilson}, {Wisniewski},
  {Wood-Vasey}, {Ye`che}, {York}, {Zakamska}, {Zamora}, {Zasowski}, {Zehavi},
  {Zhao}, {Zheng}, {Zhou}, {Zhou}, {Zou}, \& {Zhu}}]{alam15}
{Alam}, S., {Albareti}, F.~D., {Allende Prieto}, C., {et~al.} 2015, \apjs, 219,
  12

\bibitem[{{Bailer-Jones} {et~al.}(2021){Bailer-Jones}, {Rybizki}, {Fouesneau},
  {Demleitner}, \& {Andrae}}]{bailer21}
{Bailer-Jones}, C.~A.~L., {Rybizki}, J., {Fouesneau}, M., {Demleitner}, M., \&
  {Andrae}, R. 2021, \aj, 161, 147

\bibitem[{{Chatys} {et~al.}(2019){Chatys}, {Bedding}, {Murphy}, {Kiss},
  {Dobie}, \& {Grindlay}}]{chatys19}
{Chatys}, F.~W., {Bedding}, T.~R., {Murphy}, S.~J., {et~al.} 2019, \mnras, 487,
  4832

\bibitem[{{Cincunegui} {et~al.}(2007){Cincunegui}, {D{\'\i}az}, \&
  {Mauas}}]{cincunegui07}
{Cincunegui}, C., {D{\'\i}az}, R.~F., \& {Mauas}, P.~J.~D. 2007, \aap, 461,
  1107

\bibitem[{{Cutri} {et~al.}(2003){Cutri}, {Skrutskie}, {van Dyk}, {Beichman},
  {Carpenter}, {Chester}, {Cambresy}, {Evans}, {Fowler}, {Gizis}, {Howard},
  {Huchra}, {Jarrett}, {Kopan}, {Kirkpatrick}, {Light}, {Marsh}, {McCallon},
  {Schneider}, {Stiening}, {Sykes}, {Weinberg}, {Wheaton}, {Wheelock}, \&
  {Zacarias}}]{cutri03}
{Cutri}, R.~M., {Skrutskie}, M.~F., {van Dyk}, S., {et~al.} 2003, {2MASS All
  Sky Catalog of point sources.}

\bibitem[{{de Jong} {et~al.}(2012){de Jong}, {Bellido-Tirado}, {Chiappini},
  {Depagne}, {Haynes}, {Johl}, {Schnurr}, {Schwope}, {Walcher}, {Dionies},
  {Haynes}, {Kelz}, {Kitaura}, {Lamer}, {Minchev}, {M{\"u}ller}, {Nuza},
  {Olaya}, {Piffl}, {Popow}, {Steinmetz}, {Ural}, {Williams}, {Winkler},
  {Wisotzki}, {Ansorge}, {Banerji}, {Gonzalez Solares}, {Irwin}, {Kennicutt},
  {King}, {McMahon}, {Koposov}, {Parry}, {Sun}, {Walton}, {Finger}, {Iwert},
  {Krumpe}, {Lizon}, {Vincenzo}, {Amans}, {Bonifacio}, {Cohen}, {Francois},
  {Jagourel}, {Mignot}, {Royer}, {Sartoretti}, {Bender}, {Grupp}, {Hess},
  {Lang-Bardl}, {Muschielok}, {B{\"o}hringer}, {Boller}, {Bongiorno}, {Brusa},
  {Dwelly}, {Merloni}, {Nandra}, {Salvato}, {Pragt}, {Navarro}, {Gerlofsma},
  {Roelfsema}, {Dalton}, {Middleton}, {Tosh}, {Boeche}, {Caffau}, {Christlieb},
  {Grebel}, {Hansen}, {Koch}, {Ludwig}, {Quirrenbach}, {Sbordone}, {Seifert},
  {Thimm}, {Trifonov}, {Helmi}, {Trager}, {Feltzing}, {Korn}, \&
  {Boland}}]{deJong12}
{de Jong}, R.~S., {Bellido-Tirado}, O., {Chiappini}, C., {et~al.} 2012, in
  Society of Photo-Optical Instrumentation Engineers (SPIE) Conference Series,
  Vol. 8446, Ground-based and Airborne Instrumentation for Astronomy IV, ed.
  I.~S. {McLean}, S.~K. {Ramsay}, \& H.~{Takami}, 84460T

\bibitem[{{Dorda} {et~al.}(2018){Dorda}, {Negueruela}, \&
  {Gonz{\'a}lez-Fern{\'a}ndez}}]{dorda18}
{Dorda}, R., {Negueruela}, I., \& {Gonz{\'a}lez-Fern{\'a}ndez}, C. 2018,
  \mnras, 475, 2003

\bibitem[{{Drew} {et~al.}(2014){Drew}, {Gonzalez-Solares}, {Greimel}, {Irwin},
  {K{\"u}pc{\"u} Yoldas}, {Lewis}, {Barentsen}, {Eisl{\"o}ffel}, {Farnhill},
  {Martin}, {Walsh}, {Walton}, {Mohr-Smith}, {Raddi}, {Sale}, {Wright},
  {Groot}, {Barlow}, {Corradi}, {Drake}, {Fabregat}, {Frew}, {G{\"a}nsicke},
  {Knigge}, {Mampaso}, {Morris}, {Naylor}, {Parker}, {Phillipps}, {Ruhland},
  {Steeghs}, {Unruh}, {Vink}, {Wesson}, \& {Zijlstra}}]{drew14}
{Drew}, J.~E., {Gonzalez-Solares}, E., {Greimel}, R., {et~al.} 2014, \mnras,
  440, 2036

\bibitem[{{Epchtein} {et~al.}(1994){Epchtein}, {de Batz}, {Copet}, {Fouque},
  {Lacombe}, {Le Bertre}, {Mamon}, {Rouan}, {Tiphene}, {Burton}, {Deul},
  {Habing}, {Boersenberger}, {Dennefeld}, {Omont}, {Renault},
  {Rocca-Volmerange}, {Kimeswenger}, {Appenzeller}, {Bender}, {Forveille},
  {Garzon}, {Hron}, {Persi}, {Ferrari-Toniolo}, \& {Vauglin}}]{epchtein94}
{Epchtein}, N., {de Batz}, B., {Copet}, E., {et~al.} 1994, \apss, 217, 3

\bibitem[{{Feast} {et~al.}(1980){Feast}, {Catchpole}, {Carter}, \&
  {Roberts}}]{feast80}
{Feast}, M.~W., {Catchpole}, R.~M., {Carter}, B.~S., \& {Roberts}, G. 1980,
  \mnras, 193, 377

\bibitem[{{Gaia Collaboration} {et~al.}(2021){Gaia Collaboration}, {Brown},
  {Vallenari}, {Prusti}, {de Bruijne}, {Babusiaux}, {Biermann}, {Creevey},
  {Evans}, {Eyer}, {Hutton}, {Jansen}, {Jordi}, {Klioner}, {Lammers},
  {Lindegren}, {Luri}, {Mignard}, {Panem}, {Pourbaix}, {Randich}, {Sartoretti},
  {Soubiran}, {Walton}, {Arenou}, {Bailer-Jones}, {Bastian}, {Cropper},
  {Drimmel}, {Katz}, {Lattanzi}, {van Leeuwen}, {Bakker}, {Cacciari},
  {Casta{\~n}eda}, {De Angeli}, {Ducourant}, {Fabricius}, {Fouesneau},
  {Fr{\'e}mat}, {Guerra}, {Guerrier}, {Guiraud}, {Jean-Antoine Piccolo},
  {Masana}, {Messineo}, {Mowlavi}, {Nicolas}, {Nienartowicz}, {Pailler},
  {Panuzzo}, {Riclet}, {Roux}, {Seabroke}, {Sordo}, {Tanga}, {Th{\'e}venin},
  {Gracia-Abril}, {Portell}, {Teyssier}, {Altmann}, {Andrae}, {Bellas-Velidis},
  {Benson}, {Berthier}, {Blomme}, {Brugaletta}, {Burgess}, {Busso}, {Carry},
  {Cellino}, {Cheek}, {Clementini}, {Damerdji}, {Davidson}, {Delchambre},
  {Dell'Oro}, {Fern{\'a}ndez-Hern{\'a}ndez}, {Galluccio}, {Garc{\'\i}a-Lario},
  {Garcia-Reinaldos}, {Gonz{\'a}lez-N{\'u}{\~n}ez}, {Gosset}, {Haigron},
  {Halbwachs}, {Hambly}, {Harrison}, {Hatzidimitriou}, {Heiter},
  {Hern{\'a}ndez}, {Hestroffer}, {Hodgkin}, {Holl}, {Jan{\ss}en}, {Jevardat de
  Fombelle}, {Jordan}, {Krone-Martins}, {Lanzafame}, {L{\"o}ffler}, {Lorca},
  {Manteiga}, {Marchal}, {Marrese}, {Moitinho}, {Mora}, {Muinonen}, {Osborne},
  {Pancino}, {Pauwels}, {Petit}, {Recio-Blanco}, {Richards}, {Riello},
  {Rimoldini}, {Robin}, {Roegiers}, {Rybizki}, {Sarro}, {Siopis}, {Smith},
  {Sozzetti}, {Ulla}, {Utrilla}, {van Leeuwen}, {van Reeven}, {Abbas}, {Abreu
  Aramburu}, {Accart}, {Aerts}, {Aguado}, {Ajaj}, {Altavilla}, {{\'A}lvarez},
  {{\'A}lvarez Cid-Fuentes}, {Alves}, {Anderson}, {Anglada Varela}, {Antoja},
  {Audard}, {Baines}, {Baker}, {Balaguer-N{\'u}{\~n}ez}, {Balbinot}, {Balog},
  {Barache}, {Barbato}, {Barros}, {Barstow}, {Bartolom{\'e}}, {Bassilana},
  {Bauchet}, {Baudesson-Stella}, {Becciani}, {Bellazzini}, {Bernet}, {Bertone},
  {Bianchi}, {Blanco-Cuaresma}, {Boch}, {Bombrun}, {Bossini}, {Bouquillon},
  {Bragaglia}, {Bramante}, {Breedt}, {Bressan}, {Brouillet}, {Bucciarelli},
  {Burlacu}, {Busonero}, {Butkevich}, {Buzzi}, {Caffau}, {Cancelliere},
  {C{\'a}novas}, {Cantat-Gaudin}, {Carballo}, {Carlucci}, {Carnerero},
  {Carrasco}, {Casamiquela}, {Castellani}, {Castro-Ginard}, {Castro Sampol},
  {Chaoul}, {Charlot}, {Chemin}, {Chiavassa}, {Cioni}, {Comoretto}, {Cooper},
  {Cornez}, {Cowell}, {Crifo}, {Crosta}, {Crowley}, {Dafonte}, {Dapergolas},
  {David}, {David}, {de Laverny}, {De Luise}, {De March}, {De Ridder}, {de
  Souza}, {de Teodoro}, {de Torres}, {del Peloso}, {del Pozo}, {Delbo},
  {Delgado}, {Delgado}, {Delisle}, {Di Matteo}, {Diakite}, {Diener},
  {Distefano}, {Dolding}, {Eappachen}, {Edvardsson}, {Enke}, {Esquej}, {Fabre},
  {Fabrizio}, {Faigler}, {Fedorets}, {Fernique}, {Fienga}, {Figueras},
  {Fouron}, {Fragkoudi}, {Fraile}, {Franke}, {Gai}, {Garabato},
  {Garcia-Gutierrez}, {Garc{\'\i}a-Torres}, {Garofalo}, {Gavras}, {Gerlach},
  {Geyer}, {Giacobbe}, {Gilmore}, {Girona}, {Giuffrida}, {Gomel}, {Gomez},
  {Gonzalez-Santamaria}, {Gonz{\'a}lez-Vidal}, {Granvik},
  {Guti{\'e}rrez-S{\'a}nchez}, {Guy}, {Hauser}, {Haywood}, {Helmi}, {Hidalgo},
  {Hilger}, {H{\l}adczuk}, {Hobbs}, {Holland}, {Huckle}, {Jasniewicz},
  {Jonker}, {Juaristi Campillo}, {Julbe}, {Karbevska}, {Kervella}, {Khanna},
  {Kochoska}, {Kontizas}, {Kordopatis}, {Korn}, {Kostrzewa-Rutkowska},
  {Kruszy{\'n}ska}, {Lambert}, {Lanza}, {Lasne}, {Le Campion}, {Le Fustec},
  {Lebreton}, {Lebzelter}, {Leccia}, {Leclerc}, {Lecoeur-Taibi}, {Liao},
  {Licata}, {Lindstr{\o}m}, {Lister}, {Livanou}, {Lobel}, {Madrero Pardo},
  {Managau}, {Mann}, {Marchant}, {Marconi}, {Marcos Santos}, {Marinoni},
  {Marocco}, {Marshall}, {Martin Polo}, {Mart{\'\i}n-Fleitas}, {Masip},
  {Massari}, {Mastrobuono-Battisti}, {Mazeh}, {McMillan}, {Messina},
  {Michalik}, {Millar}, {Mints}, {Molina}, {Molinaro}, {Moln{\'a}r},
  {Montegriffo}, {Mor}, {Morbidelli}, {Morel}, {Morris}, {Mulone}, {Munoz},
  {Muraveva}, {Murphy}, {Musella}, {Noval}, {Ord{\'e}novic}, {Orr{\`u}},
  {Osinde}, {Pagani}, {Pagano}, {Palaversa}, {Palicio}, {Panahi}, {Pawlak},
  {Pe{\~n}alosa Esteller}, {Penttil{\"a}}, {Piersimoni}, {Pineau}, {Plachy},
  {Plum}, {Poggio}, {Poretti}, {Poujoulet}, {Pr{\v{s}}a}, {Pulone}, {Racero},
  {Ragaini}, {Rainer}, {Raiteri}, {Rambaux}, {Ramos}, {Ramos-Lerate}, {Re
  Fiorentin}, {Regibo}, {Reyl{\'e}}, {Ripepi}, {Riva}, {Rixon}, {Robichon},
  {Robin}, {Roelens}, {Rohrbasser}, {Romero-G{\'o}mez}, {Rowell}, {Royer},
  {Rybicki}, {Sadowski}, {Sagrist{\`a} Sell{\'e}s}, {Sahlmann}, {Salgado},
  {Salguero}, {Samaras}, {Sanchez Gimenez}, {Sanna}, {Santove{\~n}a},
  {Sarasso}, {Schultheis}, {Sciacca}, {Segol}, {Segovia}, {S{\'e}gransan},
  {Semeux}, {Shahaf}, {Siddiqui}, {Siebert}, {Siltala}, {Slezak}, {Smart},
  {Solano}, {Solitro}, {Souami}, {Souchay}, {Spagna}, {Spoto}, {Steele},
  {Steidelm{\"u}ller}, {Stephenson}, {S{\"u}veges}, {Szabados}, {Szegedi-Elek},
  {Taris}, {Tauran}, {Taylor}, {Teixeira}, {Thuillot}, {Tonello}, {Torra},
  {Torra}, {Turon}, {Unger}, {Vaillant}, {van Dillen}, {Vanel}, {Vecchiato},
  {Viala}, {Vicente}, {Voutsinas}, {Weiler}, {Wevers}, {Wyrzykowski}, {Yoldas},
  {Yvard}, {Zhao}, {Zorec}, {Zucker}, {Zurbach}, \& {Zwitter}}]{gaia21}
{Gaia Collaboration}, {Brown}, A.~G.~A., {Vallenari}, A., {et~al.} 2021, \aap,
  649, A1

\bibitem[{{Glass}(1979)}]{glass79}
{Glass}, I.~S. 1979, \mnras, 186, 317

\bibitem[{{Horne} \& {Baliunas}(1986)}]{horne86}
{Horne}, J.~H. \& {Baliunas}, S.~L. 1986, \apj, 302, 757

\bibitem[{{Ita} {et~al.}(2004){Ita}, {Tanab{\'e}}, {Matsunaga}, {Nakajima},
  {Nagashima}, {Nagayama}, {Kato}, {Kurita}, {Nagata}, {Sato}, {Tamura},
  {Nakaya}, \& {Nakada}}]{ita04}
{Ita}, Y., {Tanab{\'e}}, T., {Matsunaga}, N., {et~al.} 2004, \mnras, 353, 705

\bibitem[{{Ivezi{\'c}} {et~al.}(2019){Ivezi{\'c}}, {Kahn}, {Tyson}, {Abel},
  {Acosta}, {Allsman}, {Alonso}, {AlSayyad}, {Anderson}, {Andrew}, {Angel},
  {Angeli}, {Ansari}, {Antilogus}, {Araujo}, {Armstrong}, {Arndt}, {Astier},
  {Aubourg}, {Auza}, {Axelrod}, {Bard}, {Barr}, {Barrau}, {Bartlett}, {Bauer},
  {Bauman}, {Baumont}, {Bechtol}, {Bechtol}, {Becker}, {Becla}, {Beldica},
  {Bellavia}, {Bianco}, {Biswas}, {Blanc}, {Blazek}, {Blandford}, {Bloom},
  {Bogart}, {Bond}, {Booth}, {Borgland}, {Borne}, {Bosch}, {Boutigny},
  {Brackett}, {Bradshaw}, {Brandt}, {Brown}, {Bullock}, {Burchat}, {Burke},
  {Cagnoli}, {Calabrese}, {Callahan}, {Callen}, {Carlin}, {Carlson},
  {Chandrasekharan}, {Charles-Emerson}, {Chesley}, {Cheu}, {Chiang}, {Chiang},
  {Chirino}, {Chow}, {Ciardi}, {Claver}, {Cohen-Tanugi}, {Cockrum}, {Coles},
  {Connolly}, {Cook}, {Cooray}, {Covey}, {Cribbs}, {Cui}, {Cutri}, {Daly},
  {Daniel}, {Daruich}, {Daubard}, {Daues}, {Dawson}, {Delgado}, {Dellapenna},
  {de Peyster}, {de Val-Borro}, {Digel}, {Doherty}, {Dubois},
  {Dubois-Felsmann}, {Durech}, {Economou}, {Eifler}, {Eracleous}, {Emmons},
  {Fausti Neto}, {Ferguson}, {Figueroa}, {Fisher-Levine}, {Focke}, {Foss},
  {Frank}, {Freemon}, {Gangler}, {Gawiser}, {Geary}, {Gee}, {Geha}, {Gessner},
  {Gibson}, {Gilmore}, {Glanzman}, {Glick}, {Goldina}, {Goldstein}, {Goodenow},
  {Graham}, {Gressler}, {Gris}, {Guy}, {Guyonnet}, {Haller}, {Harris},
  {Hascall}, {Haupt}, {Hernandez}, {Herrmann}, {Hileman}, {Hoblitt}, {Hodgson},
  {Hogan}, {Howard}, {Huang}, {Huffer}, {Ingraham}, {Innes}, {Jacoby}, {Jain},
  {Jammes}, {Jee}, {Jenness}, {Jernigan}, {Jevremovi{\'c}}, {Johns}, {Johnson},
  {Johnson}, {Jones}, {Juramy-Gilles}, {Juri{\'c}}, {Kalirai}, {Kallivayalil},
  {Kalmbach}, {Kantor}, {Karst}, {Kasliwal}, {Kelly}, {Kessler}, {Kinnison},
  {Kirkby}, {Knox}, {Kotov}, {Krabbendam}, {Krughoff}, {Kub{\'a}nek},
  {Kuczewski}, {Kulkarni}, {Ku}, {Kurita}, {Lage}, {Lambert}, {Lange},
  {Langton}, {Le Guillou}, {Levine}, {Liang}, {Lim}, {Lintott}, {Long},
  {Lopez}, {Lotz}, {Lupton}, {Lust}, {MacArthur}, {Mahabal}, {Mandelbaum},
  {Markiewicz}, {Marsh}, {Marshall}, {Marshall}, {May}, {McKercher}, {McQueen},
  {Meyers}, {Migliore}, {Miller}, {Mills}, {Miraval}, {Moeyens}, {Moolekamp},
  {Monet}, {Moniez}, {Monkewitz}, {Montgomery}, {Morrison}, {Mueller},
  {Muller}, {Mu{\~n}oz Arancibia}, {Neill}, {Newbry}, {Nief}, {Nomerotski},
  {Nordby}, {O'Connor}, {Oliver}, {Olivier}, {Olsen}, {O'Mullane}, {Ortiz},
  {Osier}, {Owen}, {Pain}, {Palecek}, {Parejko}, {Parsons}, {Pease},
  {Peterson}, {Peterson}, {Petravick}, {Libby Petrick}, {Petry},
  {Pierfederici}, {Pietrowicz}, {Pike}, {Pinto}, {Plante}, {Plate}, {Plutchak},
  {Price}, {Prouza}, {Radeka}, {Rajagopal}, {Rasmussen}, {Regnault}, {Reil},
  {Reiss}, {Reuter}, {Ridgway}, {Riot}, {Ritz}, {Robinson}, {Roby}, {Roodman},
  {Rosing}, {Roucelle}, {Rumore}, {Russo}, {Saha}, {Sassolas}, {Schalk},
  {Schellart}, {Schindler}, {Schmidt}, {Schneider}, {Schneider}, {Schoening},
  {Schumacher}, {Schwamb}, {Sebag}, {Selvy}, {Sembroski}, {Seppala}, {Serio},
  {Serrano}, {Shaw}, {Shipsey}, {Sick}, {Silvestri}, {Slater}, {Smith},
  {Smith}, {Sobhani}, {Soldahl}, {Storrie-Lombardi}, {Stover}, {Strauss},
  {Street}, {Stubbs}, {Sullivan}, {Sweeney}, {Swinbank}, {Szalay}, {Takacs},
  {Tether}, {Thaler}, {Thayer}, {Thomas}, {Thornton}, {Thukral}, {Tice},
  {Trilling}, {Turri}, {Van Berg}, {Vanden Berk}, {Vetter}, {Virieux},
  {Vucina}, {Wahl}, {Walkowicz}, {Walsh}, {Walter}, {Wang}, {Wang}, {Warner},
  {Wiecha}, {Willman}, {Winters}, {Wittman}, {Wolff}, {Wood-Vasey}, {Wu},
  {Xin}, {Yoachim}, \& {Zhan}}]{ivezic19}
{Ivezi{\'c}}, {\v{Z}}., {Kahn}, S.~M., {Tyson}, J.~A., {et~al.} 2019, \apj,
  873, 111

\bibitem[{{Johnson}(1966)}]{johnson66}
{Johnson}, H.~L. 1966, \araa, 4, 193

\bibitem[{{Kiss} {et~al.}(2006){Kiss}, {Szab{\'o}}, \& {Bedding}}]{kiss06}
{Kiss}, L.~L., {Szab{\'o}}, G.~M., \& {Bedding}, T.~R. 2006, \mnras, 372, 1721

\bibitem[{{Koornneef}(1983)}]{koornneef83}
{Koornneef}, J. 1983, \aap, 128, 84

\bibitem[{{Landolt}(2009)}]{landolt09}
{Landolt}, A.~U. 2009, \aj, 137, 4186

\bibitem[{{Landsman}(1993)}]{landsman93}
{Landsman}, W.~B. 1993, ASPCS, ADASS II edit. by {Hanisch}, R.~J. and
  {Brissenden}, R.~J.~V. and {Barnes}, J., Vol.~52, {The IDL Astronomy User's
  Library}, 246

\bibitem[{{Messineo} \& {Brown}(2019)}]{messineo19}
{Messineo}, M. \& {Brown}, A.~G.~A. 2019, \aj, 158, 20

\bibitem[{{Messineo} \& {Brown}(2020)}]{messineo20var}
{Messineo}, M. \& {Brown}, A.~G.~A. 2020, in Stars and their Variability
  Observed from Space, ed. C.~{Neiner}, W.~W. {Weiss}, D.~{Baade}, R.~E.
  {Griffin}, C.~C. {Lovekin}, \& A.~F.~J. {Moffat}, 111--112

\bibitem[{{Messineo} \& {Brown}(2021)}]{messineo21z}
{Messineo}, M. \& {Brown}, A.~G.~A. 2021, Zenodo technical
  note,10.5281/zenodo.4964818

\bibitem[{{Messineo} {et~al.}(2021){Messineo}, {Figer}, {Kudritzki}, {Zhu},
  {Menten}, {Ivanov}, \& {Chen}}]{messineo21}
{Messineo}, M., {Figer}, D.~F., {Kudritzki}, R.-P., {et~al.} 2021, \aj, 162,
  187

\bibitem[{{Messineo} {et~al.}(2004){Messineo}, {Habing}, {Menten}, {Omont}, \&
  {Sjouwerman}}]{messineo04}
{Messineo}, M., {Habing}, H.~J., {Menten}, K.~M., {Omont}, A., \& {Sjouwerman},
  L.~O. 2004, \aap, 418, 103

\bibitem[{{Messineo} {et~al.}(2005){Messineo}, {Habing}, {Menten}, {Omont},
  {Sjouwerman}, \& {Bertoldi}}]{messineo05}
{Messineo}, M., {Habing}, H.~J., {Menten}, K.~M., {et~al.} 2005, \aap, 435, 575

\bibitem[{{Messineo} {et~al.}(2012){Messineo}, {Menten}, {Churchwell}, \&
  {Habing}}]{messineo12}
{Messineo}, M., {Menten}, K.~M., {Churchwell}, E., \& {Habing}, H. 2012, \aap,
  537, A10

\bibitem[{{Messineo} {et~al.}(2016){Messineo}, {Zhu}, {Menten}, {Ivanov},
  {Figer}, {Kudritzki}, \& {Chen}}]{messineo16}
{Messineo}, M., {Zhu}, Q., {Menten}, K.~M., {et~al.} 2016, \apjl, 822, L5

\bibitem[{{Messineo} {et~al.}(2017){Messineo}, {Zhu}, {Menten}, {Ivanov},
  {Figer}, {Kudritzki}, \& {Chen}}]{messineo17}
{Messineo}, M., {Zhu}, Q., {Menten}, K.~M., {et~al.} 2017, \apj, 836, 65

\bibitem[{{Mowlavi} {et~al.}(2021){Mowlavi}, {Rimoldini}, {Evans}, {Riello},
  {De Angeli}, {Palaversa}, {Audard}, {Eyer}, {Garcia-Lario}, {Gavras}, {Holl},
  {Jevardat de Fombelle}, {Lec{\oe}ur-Ta{\"\i}bi}, \&
  {Nienartowicz}}]{mowlavi21}
{Mowlavi}, N., {Rimoldini}, L., {Evans}, D.~W., {et~al.} 2021, \aap, 648, A44

\bibitem[{{O'Grady} {et~al.}(2021){O'Grady}, {Drout}, {Shappee}, {Bauer},
  {Fuller}, {Kochanek}, {Jayasinghe}, {Gaensler}, {Stanek}, {Holoien},
  {Prieto}, \& {Thompson}}]{oGrady21}
{O'Grady}, A.~J.~G., {Drout}, M.~R., {Shappee}, B.~J., {et~al.} 2021, in
  American Astronomical Society Meeting Abstracts, Vol.~53, American
  Astronomical Society Meeting Abstracts, 121.03

\bibitem[{{Pierce} {et~al.}(2000){Pierce}, {Jurcevic}, \&
  {Crabtree}}]{pierce00}
{Pierce}, M.~J., {Jurcevic}, J.~S., \& {Crabtree}, D. 2000, \mnras, 313, 271

\bibitem[{{Reid} \& {Goldston}(2002)}]{reid02}
{Reid}, M.~J. \& {Goldston}, J.~E. 2002, \apj, 568, 931

\bibitem[{{Scargle}(1982)}]{scargle82}
{Scargle}, J.~D. 1982, \apj, 263, 835

\bibitem[{{Sharma} {et~al.}(2022){Sharma}, {Hayden}, {Bland-Hawthorn},
  {Stello}, {Buder}, {Zinn}, {Spina}, {Kallinger}, {Asplund}, {De Silva},
  {D'Orazi}, {Freeman}, {Kos}, {Lewis}, {Lin}, {Lind}, {Martell},
  {Schlesinger}, {Simpson}, {Zucker}, {Zwitter}, {Chen}, {Cotar}, {Kafle},
  {Khanna}, {Tepper-Garcia}, {Wang}, \& {Wittenmyer}}]{sharma22}
{Sharma}, S., {Hayden}, M.~R., {Bland-Hawthorn}, J., {et~al.} 2022, \mnras,
  510, 734

\bibitem[{{Soraisam} {et~al.}(2018){Soraisam}, {Bildsten}, {Drout}, {Bauer},
  {Gilfanov}, {Kupfer}, {Laher}, {Masci}, {Prince}, {Kulkarni}, {Matheson}, \&
  {Saha}}]{soraisam18}
{Soraisam}, M.~D., {Bildsten}, L., {Drout}, M.~R., {et~al.} 2018, \apj, 859, 73

\bibitem[{{Soszy{\'n}ski} \& {Wood}(2013)}]{soszynski13}
{Soszy{\'n}ski}, I. \& {Wood}, P.~R. 2013, \apj, 763, 103

\bibitem[{{Stetson}(1987)}]{stetson87}
{Stetson}, P.~B. 1987, \pasp, 99, 191

\bibitem[{{Taniguchi} {et~al.}(2021){Taniguchi}, {Matsunaga}, {Jian},
  {Kobayashi}, {Fukue}, {Hamano}, {Ikeda}, {Kawakita}, {Kondo}, {Otsubo},
  {Sameshima}, {Takenaka}, \& {Yasui}}]{taniguchi21}
{Taniguchi}, D., {Matsunaga}, N., {Jian}, M., {et~al.} 2021, \mnras, 502, 4210

\bibitem[{{VanderPlas}(2018)}]{vanderplas18}
{VanderPlas}, J.~T. 2018, \apjs, 236, 16

\bibitem[{{Watson} {et~al.}(2006{\natexlab{a}}){Watson}, {Henden}, \&
  {Price}}]{AAVSO}
{Watson}, C.~L., {Henden}, A.~A., \& {Price}, A. 2006{\natexlab{a}}, Society
  for Astronomical Sciences Annual Symposium, 25, 47

\bibitem[{{Watson} {et~al.}(2006{\natexlab{b}}){Watson}, {Henden}, \&
  {Price}}]{watson06}
{Watson}, C.~L., {Henden}, A.~A., \& {Price}, A. 2006{\natexlab{b}}, Society
  for Astronomical Sciences Annual Symposium, 25, 47

\bibitem[{{Wood} {et~al.}(1983){Wood}, {Bessell}, \& {Fox}}]{wood83}
{Wood}, P.~R., {Bessell}, M.~S., \& {Fox}, M.~W. 1983, \apj, 272, 99

\bibitem[{{Wu} {et~al.}(2021){Wu}, {Wu}, {Zhang}, {Ren}, {Chen}, {Hsia}, {Wu},
  {Zhu}, {Li}, {Hou}, {Wang}, {Yu}, {Yu}, \& {Lamost Mrs Collaboration}}]{wu21}
{Wu}, C.-J., {Wu}, H., {Zhang}, W., {et~al.} 2021, Research in Astronomy and
  Astrophysics, 21, 096

\end{thebibliography}

\begin{ack}
An incredible amount of work was done by the CTIO/SMARTS team.
This work is dedicated to a peaceful world.
%;
This work has made use of data from the European Space Agency (ESA) 
mission {\it Gaia} (${http://www.cosmos.esa.int/gaia}$), 
processed by the {\it Gaia} Data Processing and Analysis
Consortium (DPAC, ${http://www.cosmos.esa.int/web/gaia/dpac/consortium}$). 
Funding for the DPAC has been provided by national institutions, 
in particular those institutions participating in the {\it Gaia} 
Multilateral Agreement. 
%;
This publication makes use of data products 
from the Two Micron All Sky Survey, which is a joint project of 
the University of Massachusetts and the Infrared Processing
and Analysis Center/California Institute of Technology, 
funded by the National Aeronautics and Space
Administration and the National Science Foundation. 
%;
DENIS is a joint effort of several institutes predominantly located in Europe. 
It has been supported mainly by the French Institut National des Sciences de l'Univers, 
CNRS, and French Education Ministry, the European Southern Observatory, the State of 
Baden-Wuerttemberg, and the European Commission under networks of the SCIENCE and 
Human Capital and Mobility programs, the Landessternwarte, Heidelberg and Institut d'Astrophysique de Paris. 
%;
The Pan-STARRS1 Surveys (PS1) and the PS1 public science archive have been made possible through contributions by the Institute for Astronomy, the University of Hawaii, the Pan-STARRS Project Office, the Max-Planck Society and its participating institutes, the Max Planck Institute for Astronomy, Heidelberg and the Max Planck Institute for Extraterrestrial Physics, Garching, The Johns Hopkins University, Durham University, the University of Edinburgh, the Queen's University Belfast, the Harvard-Smithsonian Center for Astrophysics, the Las Cumbres Observatory Global Telescope Network Incorporated, the National Central University of Taiwan, the Space Telescope Science Institute, the National Aeronautics and Space Administration under Grant No. NNX08AR22G issued through the Planetary Science Division of the NASA Science Mission Directorate, the National Science Foundation Grant No. AST-1238877, the University of Maryland, Eotvos Lorand University (ELTE), the Los Alamos National Laboratory, and the Gordon and Betty Moore Foundation.
%;
VPHAS+ is based on observations made with ESO Telescopes at the La Silla or Paranal Observatories under programme ID(s) 177.D-3023(B), 177.D-3023(C), 177.D-3023(D), 177.D-3023(E).
%;
This research has made use of the VizieR catalog access tool, 
CDS, Strasbourg, France, and SIMBAD database. 
This research has made use of NASA’s Astrophysics Data 
System Bibliographic Services (ADS).  
\end{ack}

\section{Funding}
This work was partially supported by the National Natural  
Science  Foundation  of  China (NSFC-11773025), 
and USTC start-up grant KY2030000054.

\end{document}